\documentclass[letterpaper]{article} % DO NOT CHANGE THIS
\usepackage{aaai2026}
\usepackage{times}  % DO NOT CHANGE THIS
\usepackage{helvet}  % DO NOT CHANGE THIS
\usepackage{courier}  % DO NOT CHANGE THIS
\usepackage[hyphens]{url}  % DO NOT CHANGE THIS
\usepackage{graphicx} % DO NOT CHANGE THIS
\usepackage{natbib}  % DO NOT CHANGE THIS AND DO NOT ADD ANY OPTIONS TO IT
\usepackage{caption} % DO NOT CHANGE THIS AND DO NOT ADD ANY OPTIONS TO IT
\usepackage{booktabs}
\usepackage{booktabs}
\usepackage{makecell}
\usepackage{multirow}
\usepackage{subcaption}
\usepackage{bm}
\usepackage{amsmath}  % align, \text 等
\usepackage{amssymb}
\usepackage{cleveref} % Added for \Cref support

\usepackage{algorithm}        % 浮动环境：\begin{algorithm}[htbp]
\usepackage{algpseudocode} 
\usepackage{xcolor}

\usepackage{threeparttable}

\usepackage{newfloat}
\usepackage{listings}
\DeclareCaptionStyle{ruled}{labelfont=normalfont,labelsep=colon,strut=off} % DO NOT CHANGE THIS
\floatstyle{ruled}
\newfloat{listing}{tb}{lst}{}
\floatname{listing}{Listing}
\nocopyright
\title{{\texttt{BDD2Seq}}: Enabling Scalable Reversible-Circuit Synthesis via Graph-to-Sequence Learning}
\author{
  Mingkai Miao\textsuperscript{\rm 1},
  Jianheng Tang\textsuperscript{\rm 2},
  Guangyu Hu\textsuperscript{\rm 2}\thanks{Corresponding authors are Hongce Zhang and Guangyu Hu.},
  Hongce Zhang\textsuperscript{\rm 1,2}\footnotemark[\value{footnote}]
}
\affiliations{
  \textsuperscript{\rm 1}Hong Kong University of Science and Technology (Guangzhou), Guangzhou, Guangdong, China\\
  \textsuperscript{\rm 2}Hong Kong University of Science and Technology, Clear Water Bay, Hong Kong\\
  mmiao815@connect.hkust-gz.edu.cn,\ \{ghuae,jtangbf\}@connect.ust.hk,\ hongcezh@ust.hk
}

\begin{document}

\maketitle

\begin{abstract}
% Binary Decision Diagrams (BDDs) are instrumental in many electronic design automation (EDA) tasks due to their compact representation of Boolean functions. BDD-based reversible circuit synthesis, particularly relevant in quantum computing, heavily depends on effective variable ordering, as the number of BDD nodes directly affects critical metrics such as Quantum Cost, a key metric reflecting the complexity and resource demands of quantum gate implementations~\cite{quantumcost}. 
% However, determining optimal variable ordering is NP-complete~\cite{NP-Complete}, and existing heuristic methods often yield suboptimal or inconsistent performance, particularly with increasing circuit complexity.
% In this paper, we propose \texttt{BDD2Seq}, a novel approach that leverages Graph Neural Networks (GNNs) and a Pointer Network Decoder enhanced by Diverse Beam Search to effectively predict promising variable orderings. By explicitly modeling circuits as graphs, our method captures structural relationships overlooked by conventional heuristics, yielding significantly more optimized BDDs and accelerated synthesis. 
% Experimental results on reversible circuit synthesis demonstrate that our method consistently achieves significant reductions in Quantum Cost and computational time compared to traditional heuristic algorithms.
% To the best of our knowledge, this is the first work to leverage graph-based learning with generative decoding strategies to address the BDD variable ordering challenge.

Binary Decision Diagrams (BDDs) are instrumental in many electronic design automation (EDA) tasks thanks to their compact representation of Boolean functions.  In BDD‑based reversible‑circuit synthesis, which is critical for quantum computing, the chosen variable ordering governs the number of BDD nodes and thus the key metrics of resource consumption, such as Quantum Cost.  Because finding an optimal variable ordering for BDDs is an NP‑complete problem, existing heuristics often degrade as circuit complexity grows. We introduce \texttt{BDD2Seq}, a graph‑to‑sequence framework that couples a Graph Neural Network encoder with a Pointer‑Network decoder and Diverse Beam Search to predict high‑quality orderings.  By treating the circuit netlist as a graph, \texttt{BDD2Seq} learns structural dependencies that conventional heuristics overlooked, yielding smaller BDDs and faster synthesis. 
% Experiments on standard ISCAS85 and LGSynth91 benchmarks—and, to stress‑test scalability, on the completely \emph{unseen} RevLib suite of reversible/quantum circuits—show that \texttt{BDD2Seq} cuts Quantum Cost and runtime by large margins over traditional methods while generalising to circuits far outside the training distribution.
Extensive experiments on three public benchmarks show that \texttt{BDD2Seq} achieves around $1.4\times$ lower Quantum Cost and $3.7\times$ faster synthesis than modern heuristic algorithms.
To the best of our knowledge, this is the first work to tackle the variable‑ordering problem in BDD‑based reversible‑circuit synthesis with a graph‑based generative model and diversity‑promoting decoding.

% To the best of our knowledge, this is the first work to tackle BDD variable ordering with a graph‑based generative model and diversity‑promoting decoding.  

\end{abstract}

% Uncomment the following to link to your code, datasets, an extended version or similar.
% You must keep this block between (not within) the abstract and the main body of the paper.
% \begin{links}
%     \link{Code}{https://github.com/tracymiao111/BDD2Seq}
%     % \link{Datasets}{https://aaai.org/example/datasets}
%     % \link{Extended version}{https://aaai.org/example/extended-version}
% \end{links}

\section{Introduction}
Binary Decision Diagrams (BDDs)~\cite{BDD}, a compact and efficient graph representation of Boolean functions, play a significant role in electronic design automation (EDA) tasks such as logic optimization, circuit synthesis and formal verification~\cite{BDD_logic_syn_veri_usage}. 
BDDs structurally represent Boolean functions through decision nodes that branch according to Boolean variables, ultimately terminating in nodes representing constant Boolean values. Reduced Ordered Binary Decision Diagrams (ROBDDs) are among the most widely adopted BDD variants due to their canonicity, ensured by adhering to a predefined variable ordering and eliminating redundant nodes through merging isomorphic subgraphs. Consequently, for a given variable ordering, the ROBDD representation of a Boolean function is unique and compact~\cite{robdd}. In the following context, we will indistinguishably use the term BDD to refer to ROBDD.\looseness=-1

BDDs have notable applications in reversible circuit synthesis, a critical area in quantum computing and low-power design~\cite{bdd_based_rev_synthesis}. BDD-based synthesis methods generally provide scalability and resource efficiency advantages, especially for handling large Boolean functions~\cite{4_rev_syn_methods}. Crucially, resource metrics such as Quantum Cost and gate count in reversible circuits directly correlate with the node count of their BDD representations~\cite{node_to_gate, bdd_based_rev_synthesis}; therefore, minimizing BDD sizes significantly improves these metrics.  However, achieving compact BDD representations strongly depends on selecting an effective variable ordering; a poor ordering can dramatically increase BDD size and compromise efficiency.

As will be detailed in the next section, existing methods for deciding variable ordering often struggle to balance computational efficiency and solution quality, exhibiting inconsistent performance especially for large-scale circuits. 
%\hongce{no need for a separate paragraph}
Due to these limitations, there is significant motivation to explore alternative methods capable of learning from past experiences and structural information inherent in circuits. Recent advances in machine learning, especially techniques that leverage structural and sequential information, provide promising pathways to address these challenges. Its success in EDA tasks, ranging from circuit design optimization to hardware formal verification, demonstrates its potential to address longstanding challenges in this domain~\cite{ML4EDA,ML4FV}. Specifically, Graph Neural Networks (GNNs) have emerged as a powerful tool for modeling and analyzing circuit structures 
as they can inherently capture the relational and structural information  in graphs and circuits can naturally be modeled as nodes and edges of graphs~\cite{Deepgate,ABGNN,TAG}.

Since GNNs do not inherently support sequential predictions, we integrate a generative Pointer Network decoder~\cite{pointernetwork}, motivated by its demonstrated success in combinatorial optimization tasks~\cite{ptr4combinatorial}. Unlike traditional deterministic methods, auto-regressive decoders produce probability distributions over candidate sequences. Leveraging this advantage, we employ Diverse Beam Search~\cite{beamsearch,DiverseBeamSearch} to simultaneously explore multiple promising candidate sequences, thus improving solution diversity and enhancing the likelihood of identifying higher-quality variable orderings.

In this work, we present a comprehensive framework \texttt{BDD2Seq} for optimizing BDD variable ordering through the integration of graph-based learning and advanced decoding strategies. Our contributions include:
\begin{itemize}
    \item \textbf{BLIF Format Circuit‑to‑Graph Representation}: We propose a novel method to embed circuit descriptions (in BLIF format) into graph-based representations. This circuit-to-graph embedding facilitates the application of GNN, enabling it to capture intricate structural and relational information inherent in circuit designs.
    % \item \textbf{Effective Graph Encoding}:
    % We implement and evaluate various graph encoding techniques, leveraging the expressive power of GNNs to generate informative embeddings that represent structural dependencies within circuit graphs.
    % \item \textbf{NLP-Inspired Sequence Generation}: To generate promising variable orderings, we employ an NLP-inspired sequence generation approach, the Pointer Network. This generative decoder is particularly suited for  variable ordering prediction, as the output is essentially a permutation of input variables.
    % \item \textbf{Enhanced Search Strategies}: 
    % We incorporate an enhanced search strategy, Diverse Beam Search, which differs from traditional NLP beam search by maintaining multiple candidate sequences throughout the search process. This approach ensures diversity among the candidates and allows them to be carried forward for subsequent tasks, improving the likelihood of finding high-quality solutions.
\item  \textbf{NLP‑Inspired Sequence Generation with Diversity}.  
We incorporate a Pointer‑Network decoder with Diverse Beam Search, which—unlike the conventional NLP beam search—retains multiple mutually dissimilar candidate orderings throughout decoding.  This diversity markedly raises the chance of discovering near‑optimal variable permutations in a single pass.
\end{itemize}

% Extensive experiments on reversible circuit synthesis benchmarks validate the effectiveness of our proposed framework, demonstrating substantial improvements over traditional methods in critical metrics, particularly Quantum Cost (QC). Notably, our approach maintains significantly lower computational complexity growth as circuit size increases, addressing a key limitation of conventional heuristics. Moreover, our method consistently delivers superior performance in terms of BDD size reduction, Quantum Cost saving and computational efficiency, particularly for large-scale circuits. 

Extensive experiments on reversible circuit synthesis tasks validate the effectiveness of our proposed framework, demonstrating substantial and consistent improvements over traditional methods in critical metrics, particularly Quantum Cost (QC). Notably, our approach maintains significantly lower computational complexity growth as circuit size increases, addressing a key limitation of conventional heuristics. 
% \texttt{BDD2Seq}\textit{ will be released after acceptance}.

% These results underscore the considerable potential of advanced graph-based learning methods and sophisticated decoding strategies for solving the BDD variable ordering problem. 
% In addition, we will open-source the framework upon acceptance of this paper.

% The remainder of this paper is organized as follows: \Cref{sec:background} introduces essential background; \Cref{sec:method} details our proposed \texttt{BDD2Seq} methodology; \Cref{sec:experiment} presents experimental results; and \Cref{sec:conclusion} concludes the paper.

\section{Background}
\label{sec:background}
\subsection{Binary Decision Diagram}
Binary Decision Diagrams (BDDs) are graph-based structures representing Boolean functions through decision nodes arranged according to Boolean variables. Reduced Ordered Binary Decision Diagrams (ROBDDs) refine BDDs by enforcing a fixed variable ordering and removing redundant nodes, achieving a canonical and compact representation.

\begin{figure}[h]
    \centering
    \includegraphics[width=1\linewidth]{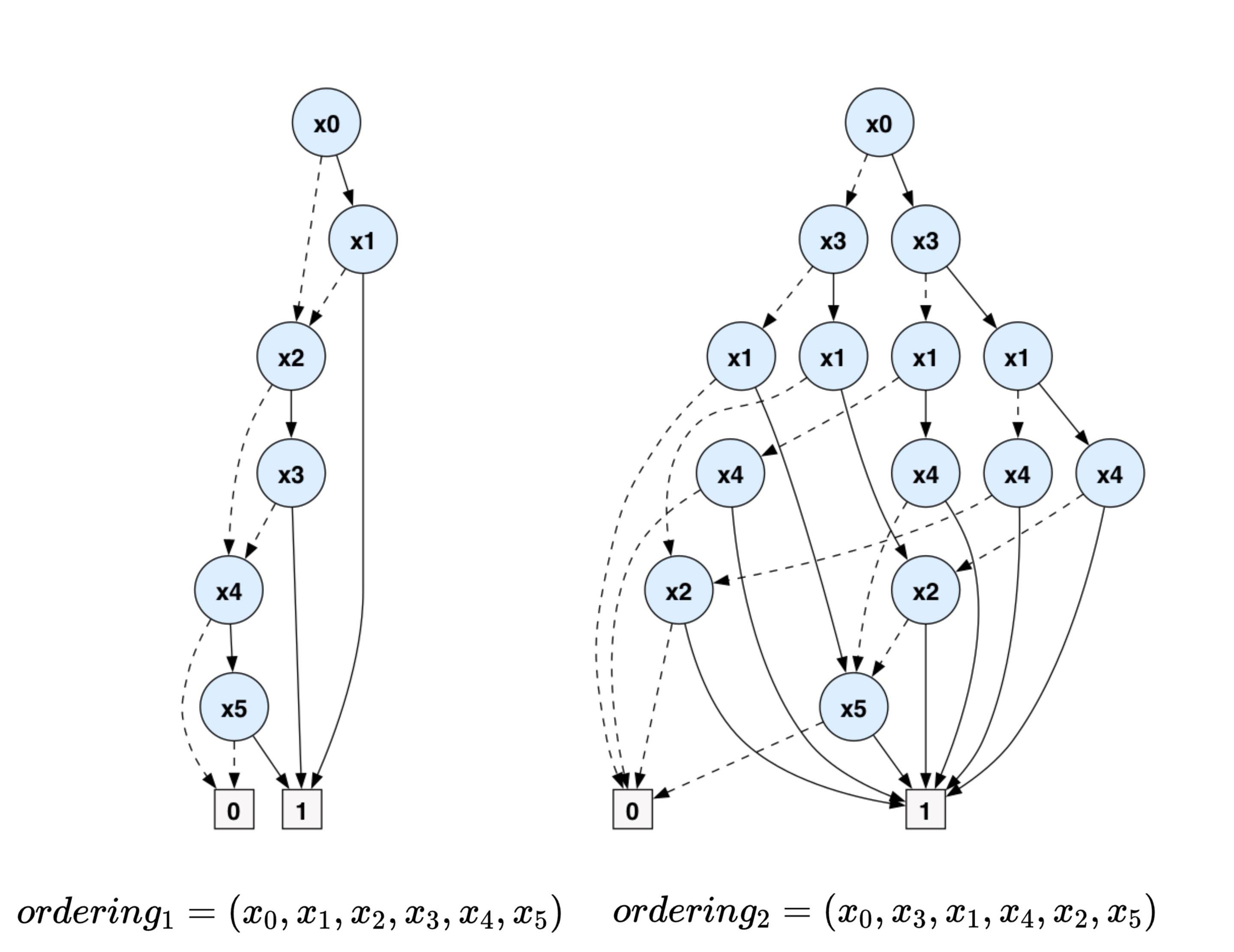}
    \caption{$f=(x_0x_1\lor x_2x_3\lor x_4x_5)$ with different orderings}
    \label{2_orderings}
\end{figure}

The size of a BDD, indicated by the node count, significantly impacts computational efficiency and resource utilization.
%in Electronic Design Automation (EDA). 
Consequently, optimizing BDD size is essential, as it directly correlates with performance metrics such as Quantum Cost, gate count, and computational resources in reversible synthesis tasks~\cite{bdd_based_rev_synthesis,node_to_gate}. One critical factor affecting BDD size is the variable ordering, which, if poorly chosen, can result in exponential growth of the BDD representation and significant slowdown of synthesis.
%significantly increasing computational complexity. 
This variability underscores the importance of selecting an effective variable ordering to maintain manageable BDD sizes. To illustrate, \textbf{\Cref{2_orderings}} shows the BDD representations of the same Boolean function \( f = (x_0 x_1 \lor x_2 x_3 \lor x_4 x_5) \) using two different variable orderings. The ordering \( (x_0, x_1, x_2, x_3, x_4, x_5) \) yields a compact BDD structure with total node count 8, whereas the ordering \( (x_0, x_3, x_1, x_4, x_2, x_5) \) leads to significant node expansion with total node count 16. This example clearly demonstrates the sensitivity of BDD node count to variable ordering and shows the importance of selecting efficient orderings to optimize performance in EDA applications.

However, identifying an optimal variable ordering is an NP-complete problem~\cite{NP-Complete}, necessitating efficient algorithms to provide near-optimal solutions. Existing approaches typically include heuristic and exact algorithms, each with distinct characteristics. Heuristic methods, such as sifting~\cite{sifting} and its variants (e.g., symmetric sifting~\cite{symm_sift} and group sifting~\cite{group_sift}), iteratively relocate variables to positions that locally minimize BDD size, prioritizing computational efficiency. Genetic algorithms (GAs)~\cite{ga} explore the search space more broadly by evolving candidate orderings through selection, crossover, and mutation. Linear methods~\cite{linear} utilize linear transformations for node-size optimization, providing rapid results with simplicity. On the other hand, exact algorithms systematically navigate the search space using techniques like dynamic programming~\cite{exact, implicit_exact_dac} to guarantee optimal solutions. 

Despite these various approaches, significant challenges persist: heuristic methods frequently converge prematurely, missing globally optimal solutions, and suffer inconsistent performance across diverse circuits, particularly for more complex Boolean functions. Conversely, exact algorithms ensure optimal solutions but are limited by their exponential computational complexity, restricting practical use to small-scale circuits. 
% Awad et al.~\cite{ga_swarm} introduced a hybrid optimizer that melds a steady-state GA with a set of modern swarm-intelligence variants, mapping their continuous moves to discrete BDD permutations and achieving near-linear scalability in both size reduction and runtime. 
A hybrid optimizer melding a steady-state genetic algorithm with several contemporary swarm-intelligence variants was introduced~\cite{ga_swarm} , translating continuous search dynamics into BDD variable permutations and delivering almost linear scalability in both size reduction and runtime.
However, its reported effectiveness is validated on only seven benchmark functions, leaving the generality of those gains open to further validation. By framing Boolean functions as hypergraphs and training a 3-hypergraph model, the method proposed in ~\cite{bdd_hefei_group} predicts near-optimal variable orderings with reduced computation time; nevertheless, its BDD node-size reduction remains inferior to both the GA and the Linear approach.

% Xu et al. ~\cite{bdd_hefei_group} framed Boolean functions as hypergraphs and trained a 3-hypergraph model to predict near-optimal variable orderings with shorter time consumption; however, its BDD node-size reduction is still weaker than GA and the Linear method mentioned above. 

% To further illustrate, in \textbf{\Cref{sifting_ga_dalu_c880_rot}}, we present a comparison of the BDD size and reordering time on three representative circuits from ISCAS85~\cite{iscas85} and LGSynth91~\cite{lgsynth91}: \texttt{c880}, \texttt{dalu}, and \texttt{rot}, 
% using the sifting algorithm---known for its efficiency---and the genetic algorithm (GA)—typically recognized for its better performance. 
To further illustrate, in \textbf{\Cref{sifting_ga_dalu_c880_rot}} we present a comparison of the BDD size and reordering time on three representative circuits from ISCAS85~\cite{iscas85} and LGSynth91~\cite{lgsynth91}: \texttt{c880}, \texttt{dalu}, and \texttt{rot}, using the sifting algorithm, known for its efficiency, and the genetic algorithm, GA, typically recognized for its better performance.
For \texttt{c880} and \texttt{dalu}, sifting is indeed quick but offers limited node-size reduction, while GA achieves improved reduction at the cost of significantly longer runtime. However, the results for the \texttt{rot} circuit reveal a critical issue: GA not only becomes excessively time-consuming but also severely underperforms in node-size reduction compared to sifting. This observation underscores a fundamental limitation in existing heuristic algorithms:  they lack the ability to learn from the intrinsic structural properties of diverse circuits, leading to inconsistent performance. 

% \begin{table}[h!]
% \centering
% % \small
% \setlength{\tabcolsep}{3pt}
% \begin{threeparttable}
%   \caption{Sifting/GA for BDD reordering on 3 samples}
%   \label{sifting_ga_dalu_c880_rot}

%   \begin{tabular}{lccccc}
%   \toprule
%   \textbf{Circuit} & \textbf{PI/PO\tnote{\dag}} & \textbf{Gates} &
%   \textbf{Algorithm} & \textbf{Time (s)} & \textbf{Nodes} \\
%   \midrule\midrule
%   \multirow{2}{*}{\texttt{c880}} & \multirow{2}{*}{60/26} & \multirow{2}{*}{383}
%     & Sifting & 0.74   & 11437 \\ \cline{4-6}
%    & & & GA & 208.03 & 6127 \\ \midrule
%   \multirow{2}{*}{\texttt{dalu}} & \multirow{2}{*}{75/16} & \multirow{2}{*}{1697}
%     & Sifting & 39.48  & 898 \\ \cline{4-6}
%    & & & GA & 63.45  & 772 \\ \midrule
%   \multirow{2}{*}{\texttt{rot}} & \multirow{2}{*}{135/107} & \multirow{2}{*}{691}
%     & Sifting & 2.73   & 9652 \\ \cline{4-6}
%    & & & GA & 240.21 & 573369 \\
%   \bottomrule
%   \end{tabular}

%   \begin{tablenotes}[flushleft]
%     \small
%     \item[\dag] PI: \emph{primary input}; PO: \emph{primary output}.
%   \end{tablenotes}
% \end{threeparttable}
% \end{table}

\begin{table}[h!]
\centering
\begin{threeparttable}
  \caption{Sifting/GA for BDD reordering on 3 samples}
  \label{sifting_ga_dalu_c880_rot}

  \begin{tabular}{@{}lccccc@{}}
  \toprule
  \textbf{Circuit} & \textbf{PI/PO\tnote{\dag}} & \textbf{Gates} &
  \textbf{Alg.\tnote{\ddag}} & \textbf{Time(s)} & \textbf{Nodes} \\
  \midrule\midrule
  \multirow{2}{*}{\texttt{c880}} & \multirow{2}{*}{60/26} & \multirow{2}{*}{383}
    & Sifting & 0.74   & 11437 \\ \cline{4-6}
   & & & GA & 208.03 & 6127 \\ \midrule
  \multirow{2}{*}{\texttt{dalu}} & \multirow{2}{*}{75/16} & \multirow{2}{*}{1697}
    & Sifting & 39.48  & 898 \\ \cline{4-6}
   & & & GA & 63.45  & 772 \\ \midrule
  \multirow{2}{*}{\texttt{rot}} & \multirow{2}{*}{135/107} & \multirow{2}{*}{691}
    & Sifting & 2.73   & 9652 \\ \cline{4-6}
   & & & GA & 240.21 & 573369 \\
  \bottomrule
  \end{tabular}

  \begin{tablenotes}[flushleft]
    \item[\dag] PI: \emph{primary input}; PO: \emph{primary output}.
    \item[{\ddag}] Alg. denotes Algorithms.

  \end{tablenotes}
\end{threeparttable}
\end{table}

Consequently, these limitations strongly motivate developing data-driven methods capable of effectively balancing computational efficiency and consistent performance of BDD size optimization.

\subsection{Reversible Circuit Synthesis}

A reversible circuit is the realization of reversible logic~\cite{reversible_computing}, where each output uniquely maps back to a specific input, ensuring that the computational process is inherently invertible. Such one-to-one mappings significantly minimize energy dissipation and power consumption, making reversible circuits particularly attractive for energy-efficient and low-power circuit designs~\cite{rev_heat}. Quantum circuits represent a prominent subclass of reversible circuits~\cite{quantum_circuit}. Particularly, arithmetic quantum circuits like quantum adders play a critical role as core subroutines in numerous significant quantum algorithms, such as Shor's factoring algorithm. Efficient quantum arithmetic circuits directly influence the feasibility and performance of quantum algorithms by reducing gate complexity, circuit depth, and resource demands, making their optimization crucial for practical quantum computing implementations~\cite{Gidney_2021}.

To efficiently synthesize such arithmetic circuits and other reversible circuits derived from classical Boolean functions, prior works have proposed BDD-based approaches for
reversible circuit synthesis~\cite{bdd_based_rev_synthesis}, which
%have proven highly effective. BDD-based reversible circuit synthesis~\cite{bdd_based_rev_synthesis} 
leverage the compactness and canonicity of BDDs to generate efficient reversible circuits, where each BDD node directly maps to a sequence of Toffoli and CNOT gates. Compared with alternative synthesis methods, BDD-based approaches typically achieve superior performance in critical metrics such as Gate Count and Quantum Cost (QC), which directly represent the resource requirements of quantum circuit implementations~\cite{4_rev_syn_methods}.\looseness=-1

While BDD-based reversible circuit synthesis offers a highly efficient method for generating reversible circuits, the challenge remains to be minimizing the number of BDD nodes. Fewer BDD nodes directly lead to a smaller, more efficient circuit with fewer gates~\cite{node_to_gate}, which ultimately reduces Quantum Cost.  \textbf{\Cref{c17_rev_table}} and \textbf{\Cref{c17_rev}} in \textbf{Appendix} explicitly illustrate this impact by comparing reversible circuits synthesized using a natural (unoptimized) ordering with those produced by optimized variable orderings. The results clearly highlight the substantial benefits achievable through effective variable ordering optimization.

% \begin{table}[h!]
% \centering
% \small
% \caption{Metrics for reversible C17 circuit with/without GA}
% \setlength{\tabcolsep}{3pt}
% \begin{tabular}{@{}lcc@{}}
% \toprule
% \textbf{Metric}           & \textbf{BDD w/o Reordering} & \textbf{w/ GA Reordering} \\ \midrule\midrule
% \textbf{Gates}            & 18                                          & \textbf{13}                                          \\
% \textbf{Lines}            & 10                                          & \textbf{9}                                           \\
% \textbf{Quantum Costs}    & 54                                          & \textbf{37}                                          \\
% \textbf{Transistor Costs} & 200                                         & \textbf{144}                                         \\ \bottomrule
% \end{tabular}

% \label{c17_rev_table}
% \end{table}

\begin{table}[h!]
\centering
\caption{Metrics for reversible C17 circuit with/without GA}
\begin{tabular}{@{}lcc@{}}
\toprule
\textbf{Metric}           & \textbf{BDD w/o Reorder} & \textbf{w/ GA Reorder} \\ \midrule\midrule
\textbf{Gates}            & 18                                          & \textbf{13}                                          \\
\textbf{Lines}            & 10                                          & \textbf{9}                                           \\
\textbf{Quantum Costs}    & 54                                          & \textbf{37}                                          \\
\textbf{Transistor Costs} & 200                                         & \textbf{144}                                         \\ \bottomrule
\end{tabular}

\label{c17_rev_table}
\end{table}

\begin{figure*}[h]
    \centering
    \includegraphics[width=1\linewidth]{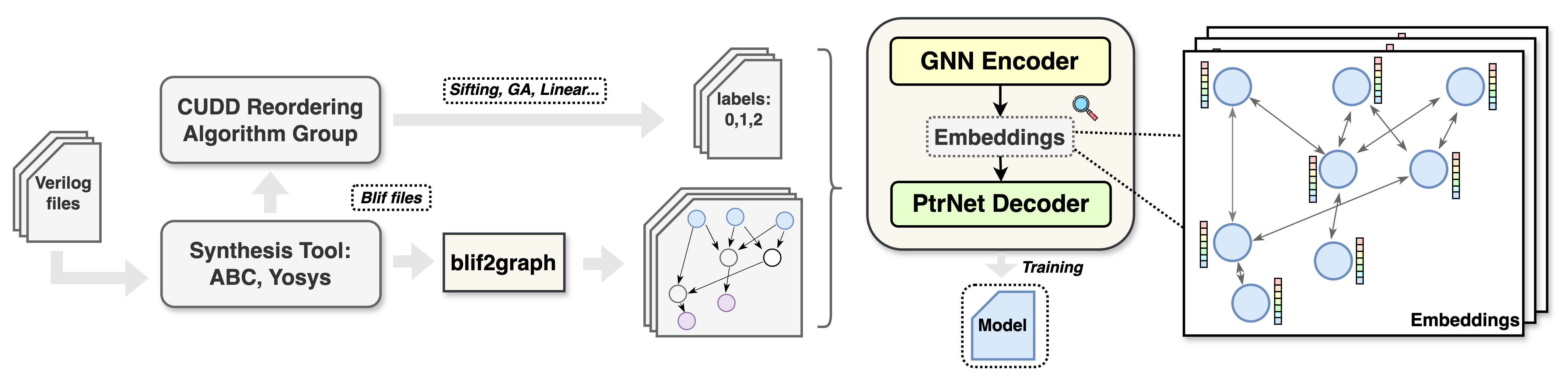}
    \caption{Overall training phase}
    \label{train}
\end{figure*}

\subsection{Graph Learning for EDA Tasks}
 Graph learning has recently attracted substantial interest within the field of EDA, primarily because many EDA tasks—including netlist optimization, circuit layout, and timing analysis—naturally map onto graph-based representations~\cite{GraphinEDA}. Unlike traditional heuristic or analytical methods, graph learning explicitly leverages structural information encoded within circuit graphs, enabling models to efficiently capture relational patterns and dependencies. This capability positions graph learning methods as promising candidates for addressing complex circuit-related optimization problems, which are challenging for conventional analytical or heuristic approaches.

 \subsection{Pointer Network with Diverse Beam Search}
A Pointer Network generates permutations by pointing to input positions, while Beam Search keeps the $k$ best partial sequences at each step~\cite{pointernetwork,beamsearch}.  
The pair already excels on permutation problems such as Traveling Salesman Problem (TSP).  
Because model likelihood and real‑world EDA objectives (e.g., Quantum Cost) do not always coincide, we couple the decoder with Diverse Beam Search~\cite{DiverseBeamSearch} to retain several mutually distinct candidates, greatly increasing the chance that at least one aligns with the downstream metric. 

\section{Methodology}\label{sec:method}

\subsection{Overview of framework}
% Overall, the framework can be broadly divided into two phases. 
In this section, we will describe the functionality of \texttt{BDD2Seq} in two phases.
The first is the training phase, which outlines the steps of data preprocessing, label generation, graph construction, and model training. The second is the inference phase, focused on the reversible circuit synthesis task, where we apply the trained model with Diverse Beam Search strategy to generate the predicted variable ordering, which is then used to perform reversible circuit synthesis.\looseness=-1

\subsubsection{Training Phase. }

\textbf{\Cref{train}} overviews the overall training phase. \emph{(i) Label generation.}  Each Verilog design is first mapped to a gate‑level BLIF netlist with ABC or Yosys~\cite{ABC,yosys}.  We then run several CUDD ordering heuristics (Sifting, GA, Linear,~\emph{etc.})~\cite{CUDD} and retain the ordering that minimises the BDD node count; this “best‑known’’ sequence serves as the supervisory label.  \emph{(ii) Graph construction.}  
The netlist is converted by \texttt{BLIF2Graph} (details given later) into a directed graph whose node features include the gate truth table and a few basic topological descriptors, such as fan-in, fan-out, and level depth, providing lightweight structural context.
% The netlist is transformed by \texttt{BLIF2Graph} (will be described later) into a directed graph whose node features encode gate truth‑tables and light topological statistics.
\emph{(iii) Model training.}  A GNN encoder embeds the graph; a Pointer‑Network decoder autoregressively outputs the variable order.  Parameters are learned end‑to‑end by minimizing a weighted negative log‑likelihood: 
\[
\mathcal{L}= \frac{1}{B}\sum_{b=1}^{B}
        \frac{\sum_{t=1}^{T}\!-\log p_\theta(y_{b,t})\,w_t\,m_{b,t}}
             {\sum_{t=1}^{T} m_{b,t}},
\]
where \(w_t\) emphasizes early positions and \(m_{b,t}\) masks padding tokens. Model parameters are optimized via backpropagation to minimize this loss.

\subsubsection{Inference Phase with Reversible Circuit Synthesis. }

\begin{figure}[h]
    \centering
    \includegraphics[width=1\linewidth]{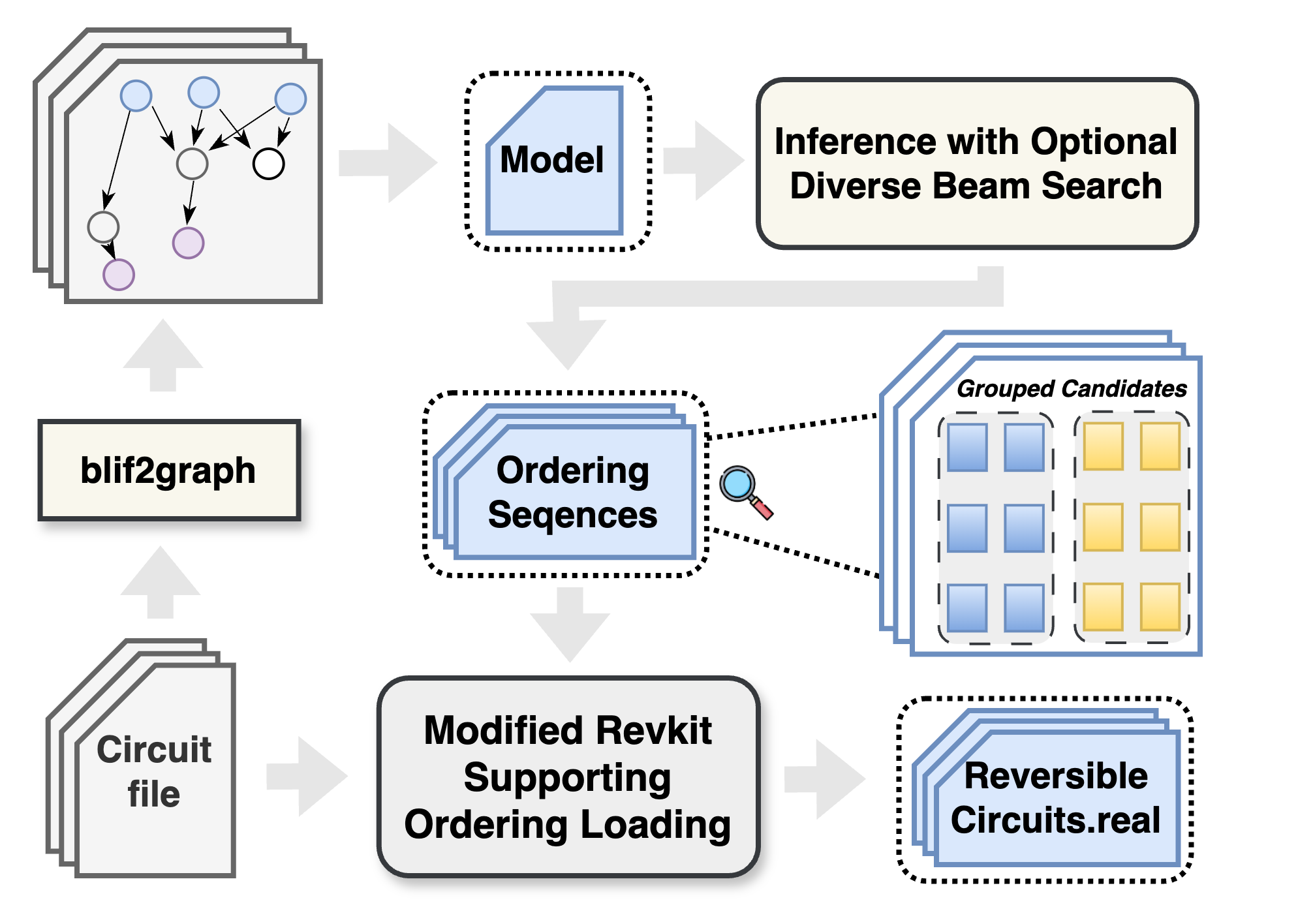}
    \caption{Inference phase with reversible circuit synthesis}
    \label{inferene}
\end{figure}

Once the model has been trained, the inference process begins by taking new circuit files ,followed by the same graph embedding step as described previously.

Shown as \textbf{\Cref{inferene}}, the core inference process involves leveraging the trained GNN encoder and Pointer Network decoder, combined with a Diverse Beam Search strategy, to predict high-quality variable ordering sequences. Diverse Beam Search enables the model to explore multiple candidate sequences, significantly increasing the likelihood of identifying an optimal or near-optimal variable ordering.

Once generated, the predicted ordering sequence is directly provided to a modified version of the Revkit synthesis framework~\cite{revkit}, a widely used reversible circuit synthesis toolkit, which we adapted to accept externally provided variable orderings. With this predicted ordering, Revkit constructs the circuit BDD and then synthesizes the reversible circuit. The synthesized circuit is subsequently outputted as a \textit{.real} file, explicitly representing the optimized reversible circuit leveraging the variable ordering predicted by our model.

% This inference process demonstrates the practical effectiveness of integrating our model with Diverse Beam Search strategy for variable ordering tasks, significantly improving solution quality and computational efficiency in reversible circuit synthesis, as will be presented in Section~\ref{sec:experiment}.

% Prior GNN-based work~\cite{aig2graph_yucunxi} converts circuits into And-Inverter-Graph (AIG/AIGER) form before building the graph representation, offering no native support for BLIF netlists, which motivates our \texttt{BLIF2Graph} embedding. 

\subsection{Circuit Embedding - \texttt{BLIF2Graph}}
\begin{figure}
    \centering
    \includegraphics[width=1\linewidth]{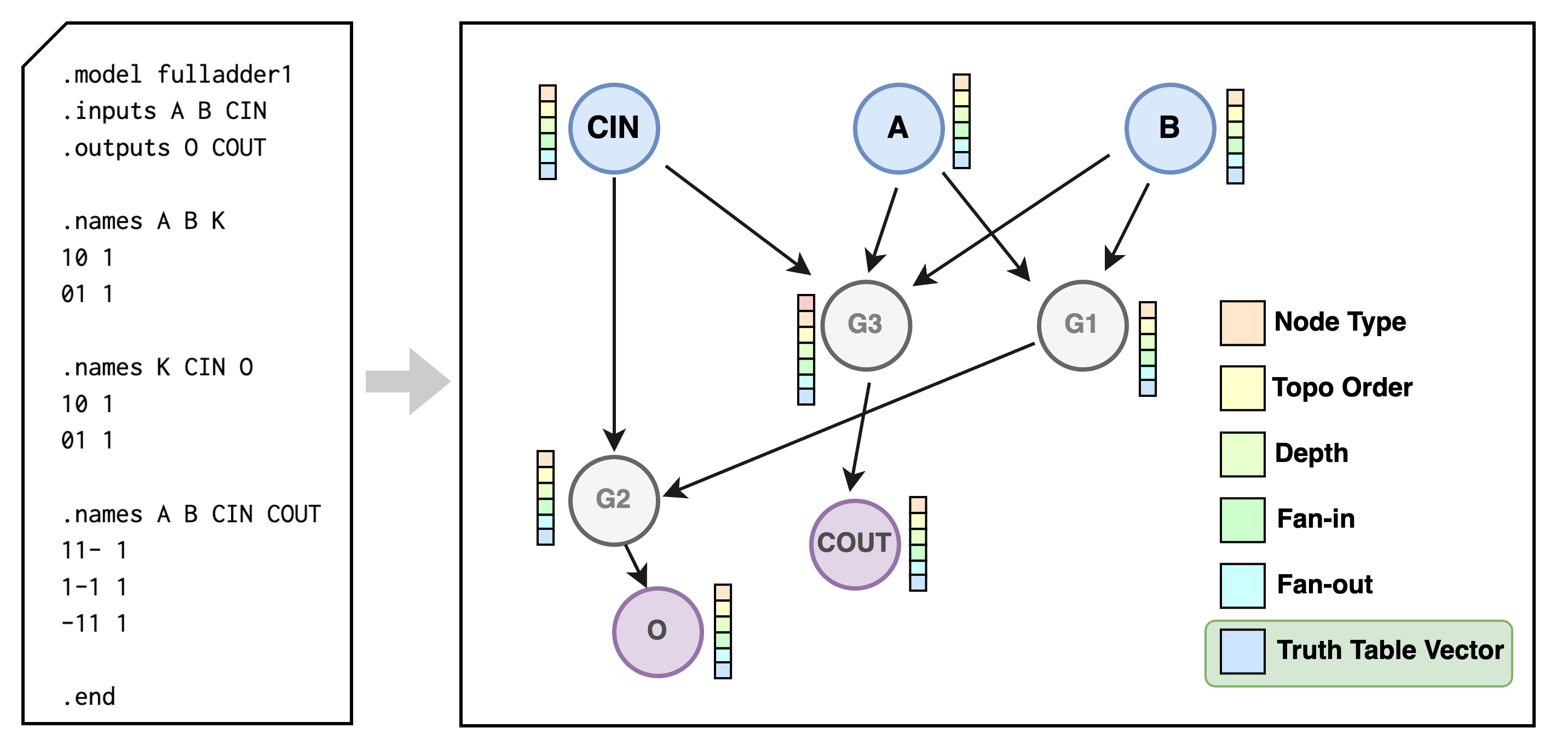}
    \caption{\texttt{BLIF2Graph}: BLIF to graph with embeddings}
    \label{blif2graph}
\end{figure}

Prior GNN-based work ~\cite{aig2graph_yucunxi} converts circuits into And Inverter Graph (AIG/AIGER) form before building the graph representation, a format that contains only AND gates and inverters and thus offers limited functional diversity; the lack of native support for BLIF netlists, where each gate is specified by a full truth table, motivates our richer \texttt{BLIF2Graph} embedding.
As illustrated in \textbf{\Cref{blif2graph}}, our proposed embedding method transforms circuit descriptions provided in BLIF format into graph representations suitable for processing by GNNs. In the BLIF format, circuits are defined through inputs, outputs, and logic gates (using \texttt{.names} as statement), where each logic gate's behavior is specified by a compact Boolean relationship between inputs and outputs.

A key innovation in our embedding method is the \textbf{truth-table encoding}, a scheme designed to enable the GNN to reason about the circuit's logic function, not just its topology. It compactly encodes the logical function of each gate into a uniform-length vector embedded directly as node features, allowing the GNN to accurately capture functional dependencies among nodes. \textbf{\Cref{truth_table_encoding}} outlines this encoding process concisely.

% Besides logical functions, we embed additional graph structural properties to comprehensively capture characteristics:

% \begin{itemize} 
% \item \textbf{Topological order}: 
% % Linear ordering respecting directed dependencies between nodes. 
% A linear arrangement of a directed acyclic graph’s nodes where each node precedes every node it has a directed edge to.
% \item \textbf{Depth}: 
% %Length of the longest path from any primary input to the node. 
% The maximum number of edges (longest directed path) from any primary input (or source) node to the node.
% \item \textbf{Fan-in/Fan-out}: Number of inputs and outputs connected to a node, indicating its connectivity within the circuit. \end{itemize}

To enrich the representation, we concatenate three lightweight structural scalars per node—{topological rank}, {depth} (longest PI‑to‑node path), and {fan‑in / fan‑out}—so the model can distinguish sources from sinks and recognize hubs without incurring extra message‑passing steps.

\begin{algorithm}[htbp]
% \caption{Truth Table Embedding for BLIF-defined Gates}
\caption{Truth Table Embedding for BLIF Gates}
\label{truth_table_encoding}
\begin{algorithmic}[1]
\Require BLIF file $B$, predefined max vector length $L$
\Ensure Uniform truth-table embeddings $V_g$ for gate $g$
\For{\textbf{each} gate $g \in B$}
    \State Extract inputs $\{x_1, x_2, ..., x_n\}$ from gate definition 
    % \State Compute $2^n$ input combinations in lexicographical order:
        \State Compute $2^n$ input combinations lexicographically:
    \[
        C_g = \left[(0 \dots 0), (0 \dots 01), \dots, (1 \dots 1)\right]
    \]
    \State Construct truth-table embedding vector $V_g$:
    \[
    V_g[i] =
    \begin{cases}
        % 1 & \text{if combination } C_i \text{ yields logic 1}\\[2pt]
        1 & \text{if } C_i \text{ yields } 1,\\[2pt]
        0 & \text{otherwise}
    \end{cases}, \quad\forall\,C_i \in C_g
    \]
    \State Pad $V_g$ to length $L$: 
    \[
        V_g \leftarrow [V_g; \mathbf{0}_{L - 2^n}]
    \]
\EndFor
\State \Return $\{V_g\}$ embeddings for all gates
\end{algorithmic}
\end{algorithm}

Integrating these structural attributes alongside logic-based truth-table embeddings enables the GNN to simultaneously model the functional and relational complexities inherent in circuit structures, thereby facilitating effective learning for optimized BDD variable ordering. We ablate graph construction by comparing \texttt{BLIF2Graph} with prior AIG-to-Graph~\cite{aig2graph_yucunxi} under identical settings; full details and results are in the \textbf{Appendix}.

\subsection{Variable Ordering Prediction}
With the circuit embeddings and GNN-based encoding presented, the next step is to decode these learned representations into variable ordering sequences. This decoding process translates graph embeddings into optimized variable permutations, directly determining the efficiency of the resulting reversible circuits.

\subsubsection{Pointer‑Network Decoding Loop. }
% Since the variable ordering task is fundamentally about generating a permutation of input variables, a Pointer Network~\cite{pointernetwork} is a natural architectural choice. Unlike traditional decoders that output to a fixed vocabulary, it directly points to input nodes, ensuring that every generated sequence is a valid permutation. \textbf{\Cref{decoding}} outlines how the Pointer Network turns node embeddings into a valid variable order.
% Decoding starts from a \texttt{<start>} token.  At each step an LSTM state~\cite{lstm} receives the previously selected node (or, under teacher forcing~\cite{teacher_forcing}, the ground‑truth node), then attends over all embeddings.  
% A dynamic mask zeros out logits of nodes already chosen, guaranteeing a permutation.  
% The masked attention scores become log‑probabilities from which the next node—and, during inference, the top‑$k$ nodes under Diverse Beam Search—is sampled.  
% The loop continues until an \texttt{<end>} token is produced, after which the collected nodes form the final ordering fed to the synthesis backend. 

\begin{figure}[!htbp] 
    \centering
    \includegraphics[width=1\linewidth]{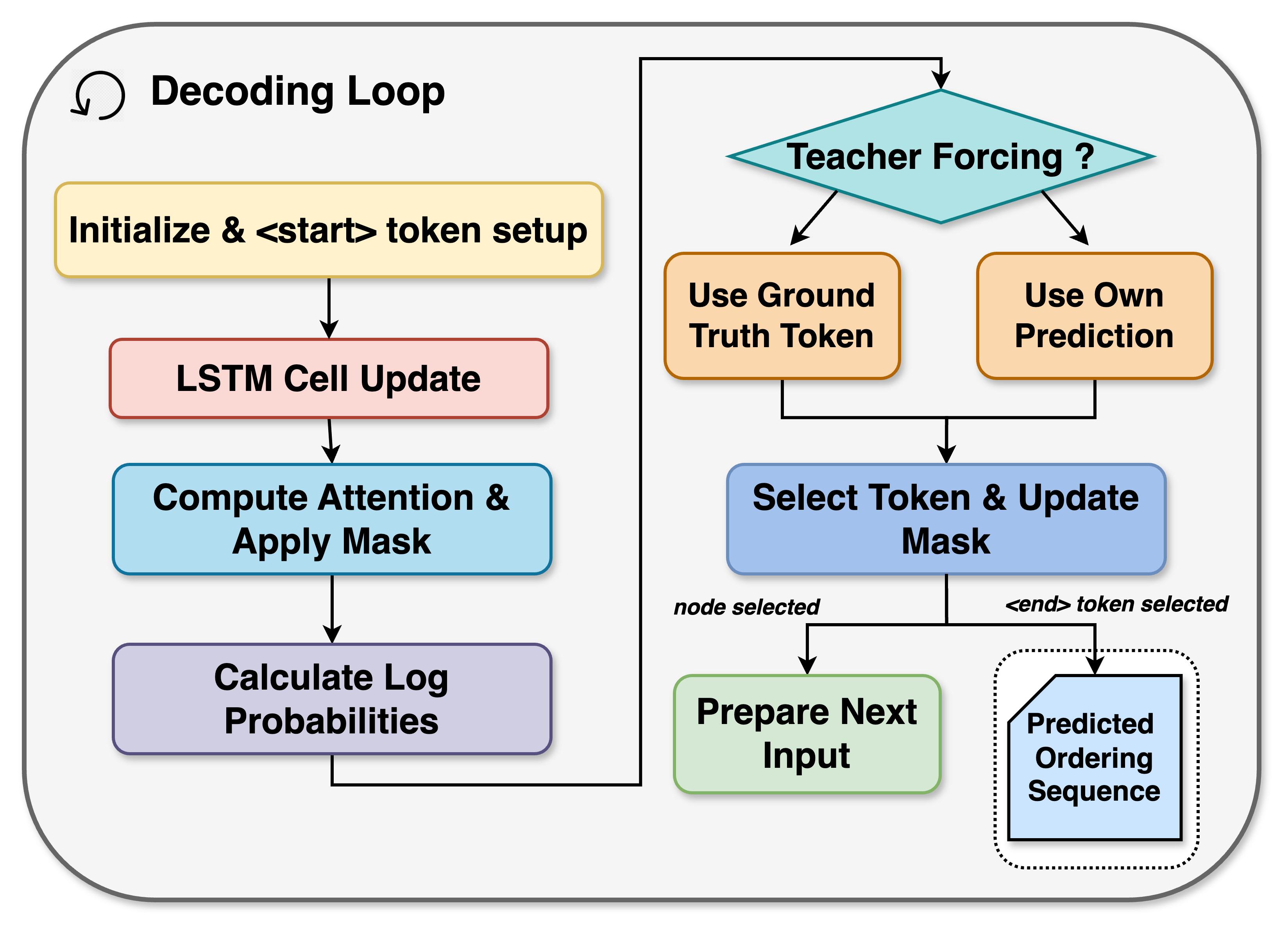}
    \caption{Pointer Network decoding loop}
    \label{decoding}

\end{figure}

Since the variable ordering problem requires producing a permutation of the input variables, a Pointer Network ~\cite{pointernetwork} is a natural choice. Rather than selecting tokens from a predefined vocabulary, the decoder points to nodes in the input graph, which guarantees that every output sequence is a valid permutation. \textbf{\Cref{decoding}} sketches the decoding loop. Decoding begins with a special \texttt{<start>} token. At each step an LSTM ~\cite{lstm} receives the previously selected variable  (or the ground-truth variable provided by teacher forcing ~\cite{teacher_forcing} during initial training), attends to all node embeddings, masks out nodes that have already been chosen, and converts the masked attention scores into log probabilities. 
% The next variable—optionally the top k under diverse beam search during inference—is sampled from this distribution. 
The decoder samples the next variable from this distribution; during inference it may instead choose the top k candidates via Diverse Beam Search.
The loop repeats until the \texttt{<end>} token is emitted, after which the collected variables constitute the final ordering passed to the synthesis backend.

\subsubsection{Diverse Beam Search. }
The likelihood produced by our model serves as a useful proxy, yet it is not perfectly correlated with the ultimate target, Quantum Cost (QC); relying solely on the single most-probable ordering can therefore be suboptimal and calls for a more diversified decoding strategy.
To raise the chance of emitting a low‑QC ordering in one forward pass, we embed \emph{Diverse Beam Search}~\cite{DiverseBeamSearch} into the decoder.

In contrast to standard beam search, Diverse Beam Search divides the beams into distinct groups, penalizing token selections repeated across these groups to encourage diversity. Formally, this diversity is enforced by applying a penalty to the raw attention scores (logits) before normalization: \begin{equation} \text{attnScores}[token] \leftarrow (1 - \alpha) \times \text{attnScores}[token], \end{equation} where $\alpha (0 \leq \alpha \leq 1)$ controls the intensity of diversity enforcement. The penalized scores are then normalized using a log-softmax operation to yield consistent log-probabilities.

\textbf{\Cref{alg:diverse_beam_search}} details this algorithm. At each decoding step, candidate beams are partitioned into \( n \) groups, each containing \(\frac{m}{n}\) sequences. Attention scores for each token are computed and penalized if previously selected by an earlier group within the current step. The highest-scoring tokens are appended, and the top sequences per group are retained for subsequent steps. This iterative process continues until all variables are ordered, producing diverse, high-quality variable ordering sequences.

By systematically exploring multiple diverse candidate orderings, our method enhances the likelihood of identifying high-quality variable orderings for the subsequent reversible circuit synthesis.

\begin{algorithm}[htbp]
\caption{Diverse Beam Search for BDD Variable Ordering Prediction}
\label{alg:diverse_beam_search}
\begin{algorithmic}[1]
\Require Beam width $m$, groups $n$, penalty $\alpha$, variables $X$
\Ensure Set of diverse variable ordering sequences
\State Initialize groups $\{G_1, \dots, G_n\}$, each with $\frac{m}{n}$ beams
\For{each decoding step $t = 1,2,...,|X|$}
    \For{each group $G_i,\; i=1,...,n$}
        \For{each candidate beam in group $G_i$}
            \State Compute raw $\text{attnScores}$ for available tokens
            \For{token previously selected in $G_j,\; j < i$}
                \[
                \text{attnScores}[{{token}]} \leftarrow (1 - \alpha) \times \text{attnScores}[{{token}]}
                \]
            \EndFor
            \State Normalize \text{attnScores} to log probabilities:
            \[
            \mathbf{p} \leftarrow \text{log\_softmax}(\text{attnScores})
            \]
            \State Append token with highest log-probability %to the sequence
        \EndFor
        \State Retain top $\frac{m}{n}$ sequences per group $G_i$
    \EndFor
\EndFor
\State \Return Final sequences from all groups
\end{algorithmic}
\end{algorithm}

\section{Experiments}\label{sec:experiment}
\subsection{Experiment Setting}
Having detailed our methodology, we now proceed to empirically validate the performance of our method on BDD-based reversible circuit synthesis.

We begin by describing the dataset preparation. We construct our dataset by data augmentation techniques including circuit decomposition, random signal negation~\cite{bdd_hefei_group}, and fuzzing via \texttt{aigfuzz}. The first two techniques take circuits from the LGSynth91 benchmark~\cite{lgsynth91} as a reference.
The resulted dataset comprises 5241 circuit variants, which are split into training, validation, and testing sets at a ratio of 7:2:1.
As for evaluation, we take the circuits from ISCAS85~\cite{iscas85}, LGSynth91, and Revlib~\cite{revlib}. 
Although the variants in training phase use LGSynth91 as reference, the original LGSynth91 circuits are unseen and are only used for evaluation purpose. Furthermore, Revlib and ISCAS85 are completely independent benchmarks.
%Note that circuits in evaluation are unseen in the training phase, especially that Revlib and ISCAS85 are completely independent.
%Additionally, circuits from ISCAS85~\cite{iscas85}, LGSynth91~\cite{lgsynth91}, and Revlib~\cite{revlib} benchmarks are used in our reversible circuit experiments to comprehensively evaluate performance.

All the experiments were conducted on a platform running Ubuntu 20.04.6 LTS, equipped with 2 NVIDIA GeForce RTX~3090 GPUs and dual Intel\textsuperscript{\textregistered} Xeon\textsuperscript{\textregistered} Platinum~8375C CPUs at 2.90\,GHz.

\subsection{Graph Encoder Selection}
To identify an effective graph encoder for predicting BDD variable orderings, we conducted a comparative evaluation of four prominent GNN architectures: Graph Convolutional Network (GCN)~\cite{GCN}, Graph Isomorphism Network (GIN)~\cite{GIN}, GraphSAGE~\cite{graphsage}, and Graph Attention Network (GAT)~\cite{GAT}.
Ranking fidelity was measured with Kendall’s~$\tau$ and Spearman’s~$\rho$,
\[
\tau=\frac{C-D}{\tfrac12 n(n-1)}, \qquad
\rho=1-\frac{6\sum_i d_i^{2}}{n(n^{2}-1)},
\]
where $C$ and $D$ count concordant and discordant pairs, $d_i$ is the rank difference for element $i$, and $n$ is the sequence length. As shown in \textbf{\Cref{gnn_metrics}} in \textbf{Appendix}, GAT achieves the highest test‑set correlation ($\tau=0.6182$, $\rho=0.6748$) and is thus adopted as our default encoder. Given the task difficulty, with some instances involving permutations of exceeding 200 variables, achieving these scores constitutes a strong rank correlation.

\begin{table*}[h!]

\centering
\begin{threeparttable}
\caption{Reversible circuit synthesis metrics comparison using the evaluation dataset}
\begin{tabular}{@{}lccccccccc@{}}
\toprule
\textbf{Metric}  & \textbf{Quality (50$^*$)}         & \textbf{Balance (20$^*$)} & \textbf{Efficiency} & \textbf{SIFT} & \textbf{SYMM$^\dag$} & \textbf{GROUP$^\dag$} & \textbf{GA} & \textbf{LINEAR} \\ \midrule\midrule
Gates         &\textbf{96643}           & \textbf{98437}                & \textbf{108181} & 144112           & 135112   & 161648  & 2547934  & 134377  \\ 
Lines      &     \textbf{28962}         & \textbf{29462}             & \textbf{31968}  & 41213           & 38876   & 45602   & 686839   & 38744  \\ 
Quantum Cost     &\textbf{307367}       & \textbf{312345}            &  \textbf{346425}  & 471792          & 439452 & 532588   & 8628394  & 437213   \\ 
Trans. Cost\ddag     &\textbf{1164728}    & \textbf{1184520}           &  \textbf{1309936}  & 1776136        & 1657200 & 2001736 & 32346744  & 1648568    \\ 
Time (s)   &  2076.2   & \textbf{521.1}           &   \textbf{156.2} & 576.7        & 630.2 & 600.4 & 735.8  & 581.0 \\ \bottomrule

\end{tabular}
\begin{tablenotes}[flushright]
\item[*] The numbers in parentheses indicate beam width, i.e., Quality (50) means beam width = 50 for Quality mode. 
\item[\dag] SYMM: symmetric sifting; GROUP: group sifting.
\item[\ddag] Trans. Cost denotes the transistor cost.
\end{tablenotes}
\label{Three_Mode_Performance}
\end{threeparttable}
\end{table*}

\begin{table*}[h!]
    \centering
    \begin{threeparttable}
    \caption{Quantum Cost comparison: \texttt{BDD2Seq} vs. traditional algorithms\textsection}
    \label{tab:all-results-qc}
    \begin{tabular}{@{}l c |c|c| c| c| c| c| c@{}}
    \toprule
    \multirow{2}{*}{\textbf{Circuit}\ddag} & \multirow{2}{*}{\textbf{PI/PO}} & \multicolumn{2}{c|}{\texttt{\textbf{BDD2Seq}} (QC*)} & \multicolumn{5}{c}{\textbf{Heuristics} (QC*)} \\
    & & \textbf{Balance} & \textbf{Efficiency} & \textbf{SIFT} & \textbf{SYMM}\dag & \textbf{GROUP}\dag & \textbf{GA} & \textbf{LINEAR} \\
    \midrule
    \texttt{dc1\_142}\textsuperscript{R} & 4/7 & \textbf{160} & 168 & 160 & 186 & 186 & 186 & 181\\
    \texttt{c17}\textsuperscript{I} & 5/2 & \textbf{37} & 49 & 49 & 49 & 37 & 37 & 49\\
    \texttt{bw\_116}\textsuperscript{R} & 5/28 & \textbf{924} & \textbf{937} & 943 & 937 & 943 & 937 & 943\\
    \texttt{con1\_136}\textsuperscript{R} & 7/2 & \textbf{88} & 103 & 96 & 95 & 96 & 95 & 96\\
    \texttt{inc\_170}\textsuperscript{R} & 7/9 & \textbf{579} & \textbf{579} & 579 & 592 & 592 & 592 & 621\\
    \texttt{alu2\_96}\textsuperscript{R} & 10/6 & \textbf{1218} & 1415 & 1436 & 1299 & 1366 & 1298 & 1376\\
    \texttt{cm151a}\textsuperscript{L} & 12/2 & \textbf{70} & \textbf{70} & 92 & 92 & 92 & 92 & 92\\
    \texttt{add6\_92}\textsuperscript{R} & 12/7 & \textbf{118} & 566 & 499 & 118 & 474 & 118 & 499\\
    \texttt{alu4\_98}\textsuperscript{R} & 14/8 & 4455 & 4934 & 7222 & \textbf{4334} & 7623 & 4403 & 5583\\
    \texttt{t481\_208}\textsuperscript{R} & 16/1 & \textbf{139} & \textbf{140} & 152 & 152 & 152 & 152 & 152\\
    \texttt{pm1}\textsuperscript{L} & 16/13 & \textbf{234} & 273 & 273 & 244 & 267 & 244 & 261\\
    \texttt{vda}\textsuperscript{L} & 17/39 & 4224 & 4224 & 4477 & 4477 & 4460 & \textbf{4169} & 4470\\
    \texttt{mux}\textsuperscript{L} & 21/1 & \textbf{135} & \textbf{135} & 170 & 170 & 170 & 170 & 170\\
    \texttt{cm150a\_128}\textsuperscript{R} & 21/1 & \textbf{136} & \textbf{136} & 186 & 186 & 186 & 186 & 186\\
    \texttt{frg1\_160}\textsuperscript{R} & 28/3 & \textbf{598} & 629 & 747 & 599 & 827 & 653 & 747\\
    \texttt{c880}\textsuperscript{I} & 60/26 & 37802 & 39992 & 61131 & 61131 & 78157 & \textbf{35122} & 61216\\
    \texttt{dalu}\textsuperscript{L} & 75/16 & 5297 & 5652 & 31145 & 31145 & 47170 & \textbf{5235} & 14330\\
    \texttt{x4}\textsuperscript{L} & 94/71 & \textbf{2662} & \textbf{2782} & 4459 & 4459 & 4334 & 3006 & 4459\\
    \texttt{apex5\_104}\textsuperscript{R} & 117/88 & 9852 & 10697 & 10349 & \textbf{9467} & 10227 & 10283 & 10349\\
    \texttt{rot}\textsuperscript{L} & 135/107 & \textbf{38453} & \textbf{48344} & 78639 & 78375 & 110887 & 7339159 & 70456\\
    \texttt{frg2}\textsuperscript{L} & 143/139 & \textbf{7189} & \textbf{7189} & 12468 & 12361 & 12154 & 9391 & 12111\\
    \texttt{pair}\textsuperscript{L} & 173/137 & \textbf{20754} & \textbf{20799} & 46544 & 46444 & 49917 & 1039245 & 47818\\
    \midrule
    Total &  & \textbf{135124} & \textbf{149813} & 261816 & 256912 & 330317 & 8454773 & 236165\\
    \bottomrule
    \end{tabular}
    \begin{tablenotes}
    \item[*] Metric: Quantum Cost (QC).
    \item[\dag] SYMM: symmetric sifting; GROUP: group sifting.
    \item[\ddag] Superscripts denote datasets: \textsuperscript{I} ISCAS85, \textsuperscript{L} LGSynth91, \textsuperscript{R} Revlib.
    \item[\textsection] See Table~\ref{tab:all-results1} and Table~\ref{tab:all-results2} in the \textbf{Appendix} for detailed results.
    \end{tablenotes}
    \end{threeparttable}
\end{table*}

\begin{table*}[h!]
    \centering
    \begin{threeparttable}
    \caption{Time consumption (seconds): \texttt{BDD2Seq} vs. traditional algorithms\textsection}
    \label{tab:all-results-time}
    \begin{tabular}{@{}l |c |c|c |c| c| c| c| c@{}}
    \toprule
    \multirow{2}{*}{\textbf{Circuit}\ddag} & \multirow{2}{*}{\textbf{PI/PO}} & \multicolumn{2}{c|}{\texttt{\textbf{BDD2Seq}} (s)} & \multicolumn{5}{c}{\textbf{Heuristics} (s)} \\
    & & \textbf{Balance} & \textbf{Efficiency} & \textbf{SIFT} & \textbf{SYMM}\dag & \textbf{GROUP}\dag & \textbf{GA} & \textbf{LINEAR} \\
    \midrule
    \texttt{dc1\_142}\textsuperscript{R} & 4/7 & 0.07 & \textbf{0.01} & 0.01 & 0.05 & 0.06 & 0.01 & 0.05\\
    \texttt{c17}\textsuperscript{I} & 5/2 & 0.08 & \textbf{0.01} & 0.07 & 0.05 & 0.06 & 0.01 & 0.01\\
    \texttt{bw\_116}\textsuperscript{R} & 5/28 & 0.08 & 0.02 & \textbf{0.01} & 0.09 & 0.07 & 0.01 & 0.06\\
    \texttt{con1\_136}\textsuperscript{R} & 7/2 & 0.13 & 0.02 & \textbf{0.01} & 0.06 & 0.05 & 0.01 & 0.05\\
    \texttt{inc\_170}\textsuperscript{R} & 7/9 & 0.13 & \textbf{0.01} & 0.01 & 0.08 & 0.07 & 0.01 & 0.06\\
    \texttt{alu2\_96}\textsuperscript{R} & 10/6 & 0.23 & 0.03 & \textbf{0.01} & 0.15 & 0.08 & 0.03 & 0.08\\
    \texttt{cm151a}\textsuperscript{L} & 12/2 & 0.29 & 0.02 & 0.06 & 0.05 & 0.05 & 0.02 & \textbf{0.01}\\
    \texttt{add6\_92}\textsuperscript{R} & 12/7 & 0.29 & \textbf{0.01} & 0.01 & 0.12 & 0.08 & 0.02 & 0.08\\
    \texttt{alu4\_98}\textsuperscript{R} & 14/8 & 0.44 & 0.09 & \textbf{0.03} & 0.65 & 0.20 & 0.18 & 0.17\\
    \texttt{t481\_208}\textsuperscript{R} & 16/1 & 0.47 & 0.03 & \textbf{0.01} & 0.11 & 0.08 & 0.04 & 0.08\\
    \texttt{pm1}\textsuperscript{L} & 16/13 & 0.46 & 0.02 & 0.06 & 0.05 & 0.06 & 0.02 & \textbf{0.01}\\
    \texttt{vda}\textsuperscript{L} & 17/39 & 0.56 & 0.08 & 0.14 & 0.14 & 0.15 & 0.19 & \textbf{0.03}\\
    \texttt{mux}\textsuperscript{L} & 21/1 & 1.03 & 0.39 & 0.24 & \textbf{0.23} & 0.26 & 0.51 & 0.43\\
    \texttt{cm150a\_128}\textsuperscript{R} & 21/1 & 0.84 & 0.24 & 0.25 & 0.25 & \textbf{0.19} & 0.33 & 0.37\\
    \texttt{frg1\_160}\textsuperscript{R} & 28/3 & 1.34 & 0.03 & \textbf{0.01} & 0.25 & 0.08 & 0.20 & 0.08\\
    \texttt{c880}\textsuperscript{I} & 60/26 & 26.22 & \textbf{15.52} & 47.71 & 44.43 & 45.20 & 121.45 & 16.86\\
    \texttt{dalu}\textsuperscript{L} & 75/16 & \textbf{67.32} & \textbf{28.17} & 182.95 & 212.63 & 202.55 & 270.63 & 431.41\\
    \texttt{x4}\textsuperscript{L} & 94/71 & 11.84 & 0.11 & 0.19 & 0.20 & 0.20 & 0.64 & \textbf{0.04}\\
    \texttt{apex5\_104}\textsuperscript{R} & 117/88 & 18.27 & 0.20 & \textbf{0.08} & 3.62 & 0.41 & 1.47 & 0.42\\
    \texttt{rot}\textsuperscript{L} & 135/107 & 47.58 & \textbf{15.69} & 61.42 & 60.18 & 61.68 & 139.15 & 20.52\\
    \texttt{frg2}\textsuperscript{L} & 143/139 & 27.54 & 0.35 & 1.19 & 1.20 & 1.18 & 2.41 & \textbf{0.25}\\
    \texttt{pair}\textsuperscript{L} & 173/137 & 39.87 & 1.06 & 4.86 & 4.81 & 4.91 & 18.63 & \textbf{1.03}\\
    \midrule
    Total &  & \textbf{245.08} & \textbf{62.15} & 299.32 & 329.39 & 317.67 & 555.94 & 472.03\\
    \bottomrule
    \end{tabular}
    \begin{tablenotes}
    \item[\ddag] Superscripts denote datasets: \textsuperscript{I} ISCAS85, \textsuperscript{L} LGSynth91, \textsuperscript{R} Revlib.
    \item[\dag] SYMM: symmetric sifting; GROUP: group sifting.
    \item[\textsection] Synthesis time reported as 0.01$s$ represents Revkit's measurement limit ($\leq 0.01s$).
    
    \end{tablenotes}
    \end{threeparttable}
\end{table*}

\subsection{Evaluation on Reversible Circuit Synthesis}
% Having assessed \texttt{BDD2Seq}'s effectiveness in reducing BDD size and computational efficiency, 
We now evaluate the practical impact on reversible circuit synthesis. Experiments are conducted on 148 combinational circuits from ISCAS85, LGSynth91, and Revlib.
% , utilizing a modified version of Revkit that supports external variable orderings.

% \subsubsection{Ablation Study for Diverse Beam Search.}

% As mentioned, to enhance the search capability for potential optimal solutions, we adapt the Diverse Beam Search into our framework. However, generating multiple candidate sequences inevitably increases computational time during reversible circuit synthesis. Therefore, to carefully balance solution quality and computational efficiency, we conducted a detailed analysis of how varying the beam width and diversity penalty parameters impacts performance.

% As illustrated in \textbf{\Cref{beam_width_penalty_new}}, we initially set the number of beam groups to half of the beam width and tested beam widths of 5, 10, 20, 30, and 50. We found that a beam width of 20 offers the most favorable balance, as further increases resulted in marginal performance improvement but significantly higher computational time. Subsequent, with the beam width fixed at 20, we tested different penalty parameters (0.1, 0.2, 0.25, 0.3, and 0.4). The results indicate that a penalty of 0.25 optimally balanced diversity and Quantum Cost reduction.

\subsubsection{Ablation Study on Diverse Beam Search. }
\begin{figure}[!htbp] 
    \centering
    \includegraphics[width=1\linewidth]{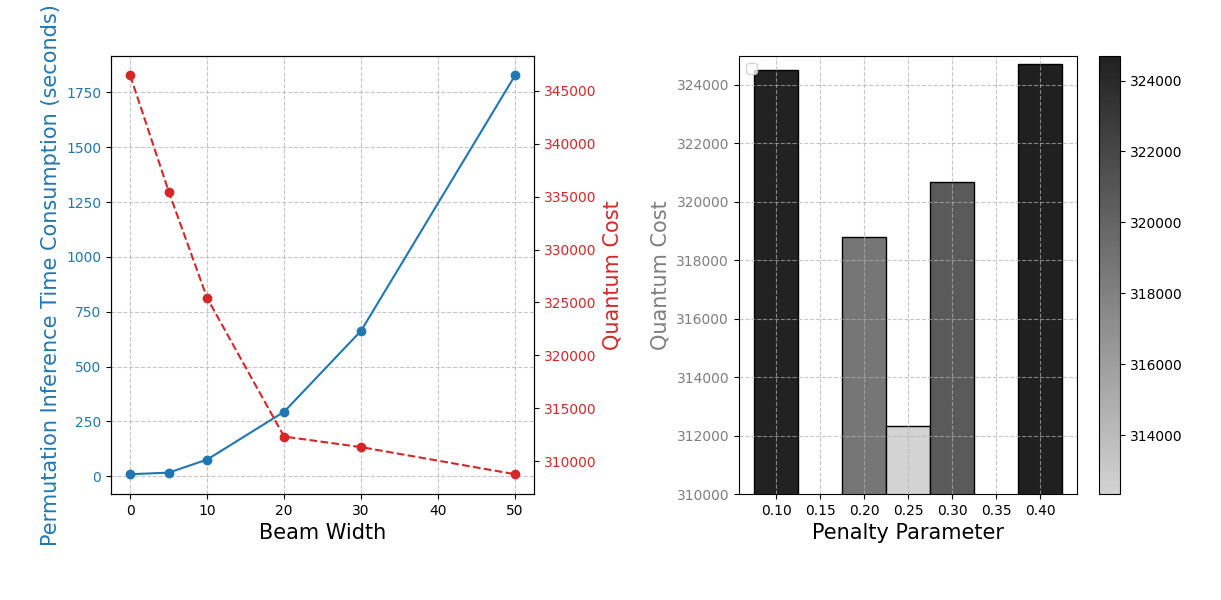}
    % \caption{Impact of beam width and penalty parameter on Quantum Cost and time consumption}
    \caption{Ablation experiment on Diverse Beam Search}

    \label{beam_width_penalty_new}

\end{figure}
To quantify the benefit of adapting Diverse Beam Search and to identify cost‑effective settings, we ablate the two hyper‑parameters: beam width~$m$ and diversity penalty~$\alpha$ as shown in~\textbf{\Cref{beam_width_penalty_new}}.  All other components of \texttt{BDD2Seq} are held fixed. Using $m/2$ diversity groups, increasing $m$ from 0 to 20 cuts Quantum Cost monotonically, confirming that Diverse Beam Search indeed surfaces better orderings.  Beyond $m{=}20$ the quality gain saturates while runtime rises sharply, so we adopt \emph{$m{=}20$} as the best cost-speed trade‑off.  With $m{=}20$, sweeping $\alpha$ shows a bowl‑shaped curve: weak penalties ($\alpha{<}0.2$) yield insufficient sequence variety, whereas overly strong penalties ($\alpha{>}0.3$) scatter probability mass onto low‑quality beams.  The minimum Quantum Cost occurs at \emph{$\alpha{=}0.25$}, which balances exploration and exploitation. These empirically chosen values ($m{=}20$, $\alpha{=}0.25$) define the \textbf{Balance} mode used later. The ablation confirms that Diverse Beam Search is indispensable for our performance gains and that its impact can be tuned without prohibitive runtime overhead.

\subsubsection{Overall Reversible‑Synthesis Comparison. }
Thanks to the tunable beam width in Diverse Beam Search, \texttt{BDD2Seq} can be deployed in three practical modes.  
\textbf{Efficiency} (greedy decoding) sacrifices some optimality for minimum latency; \textbf{Balance} fixes the beam width at~20 and targets a cost–speed sweet spot; \textbf{Quality} widens the beam to~50 to chase the lowest Quantum Cost.

\textbf{\Cref{Three_Mode_Performance}} shows that, regardless of modes, \texttt{BDD2Seq} beats classical heuristics on every metric.  The Balance configuration is usually preferable: it trims Quantum Cost by $1.4\times$–$27.6\times$ while keeping runtime moderate.  When turnaround time dominates, the Efficiency mode is still $3.7$–$4.7\times$ faster than heuristic re‑ordering yet continues to lower cost.  At the other extreme, the Quality mode pushes cost even lower, albeit with proportionally longer runtime.

\subsubsection{Detailed Evaluation on Benchmark Circuits. }
\textbf{\Cref{tab:all-results-qc}} and \textbf{\Cref{tab:all-results-time}} contrast \texttt{BDD2Seq} with mainstream heuristics over \mbox{22} benchmarks spanning from 4 to 173 primary inputs, reflecting the complexity of the search space. On most of these benchmarks, \texttt{BDD2Seq} delivers the lowest or tied-lowest Quantum Cost. In addition, runtime grows far more slowly than that of heuristic re-ordering, making our approach both cheaper and faster as circuit complexity rises.

Between the two operating modes, \textbf{Balance} (beam width $20$) secures the best or tied-best Quantum Cost on \mbox{17/22} circuits while incurring only a modest runtime increase over the ultra-fast \textbf{Efficiency} mode. In addition, although GA and Linear heuristics are reported to be two of the strongest variable-ordering techniques~\cite{bdd_hefei_group}, \texttt{BDD2Seq} surpasses both, delivering lower Quantum Cost and shorter synthesis time.

% These results confirm that \texttt{BDD2Seq} sustains high solution quality and superior scalability across a wide spectrum of reversible‑synthesis tasks.

\section{Conclusion}\label{sec:conclusion}
In this work, we introduced \texttt{BDD2Seq}, a novel deep learning approach to address the problem of optimal variable ordering in BDDs specifically for reversible circuit synthesis. 
From graph to sequence learning with enhanced searching strategy, our method effectively captures circuit structural and functional information and efficiently explores promising variable permutations.
% Our method combines a Graph Attention Network encoder, a Pointer Network decoder, and Diverse Beam Search, effectively capturing circuit structural information and efficiently exploring promising variable permutations.

Extensive experimental results confirm that \texttt{BDD2Seq} significantly outperforms traditional heuristic methods, achieving notable reductions in Quantum Cost and substantially faster synthesis time, particularly in circuits with large numbers of primary inputs. 
While its three operating modes let users trade runtime for additional cost savings when needed.
% The flexible operation modes of \texttt{BDD2Seq}—\textbf{Efficiency}, \textbf{Balance} and \textbf{Quality}—provide users the ability to tailor synthesis outcomes between speed and quality.

\section*{Acknowledgments}
We thank Yusheng Zhao and Mingfei Yu for their helpful discussions and support for this work.
\bibliography{aaai2026}

\begingroup        

\def\isChecklistMainFile{}  

\endgroup

\section{Appendix}
\subsection{Schematics of Reversible C17 Circuit with or without BDD Variable Reordering}
\textbf{\Cref{c17_rev}} visually compares the reversible implementation of the C17 benchmark under two conditions: (a) without BDD variable-ordering optimization and (b) with an optimized ordering obtained by the genetic algorithm.  The side-by-side schematics highlight how an appropriate reordering of decision-diagram variables translates into a noticeably more compact reversible realization, with shorter cascades of Toffoli controls and fewer ancillary interactions.  
\begin{figure}[h!]
    \centering
    \begin{subfigure}{0.23\textwidth}  % Set the width of the subfigure
        \centering
        \includegraphics[width=\linewidth]{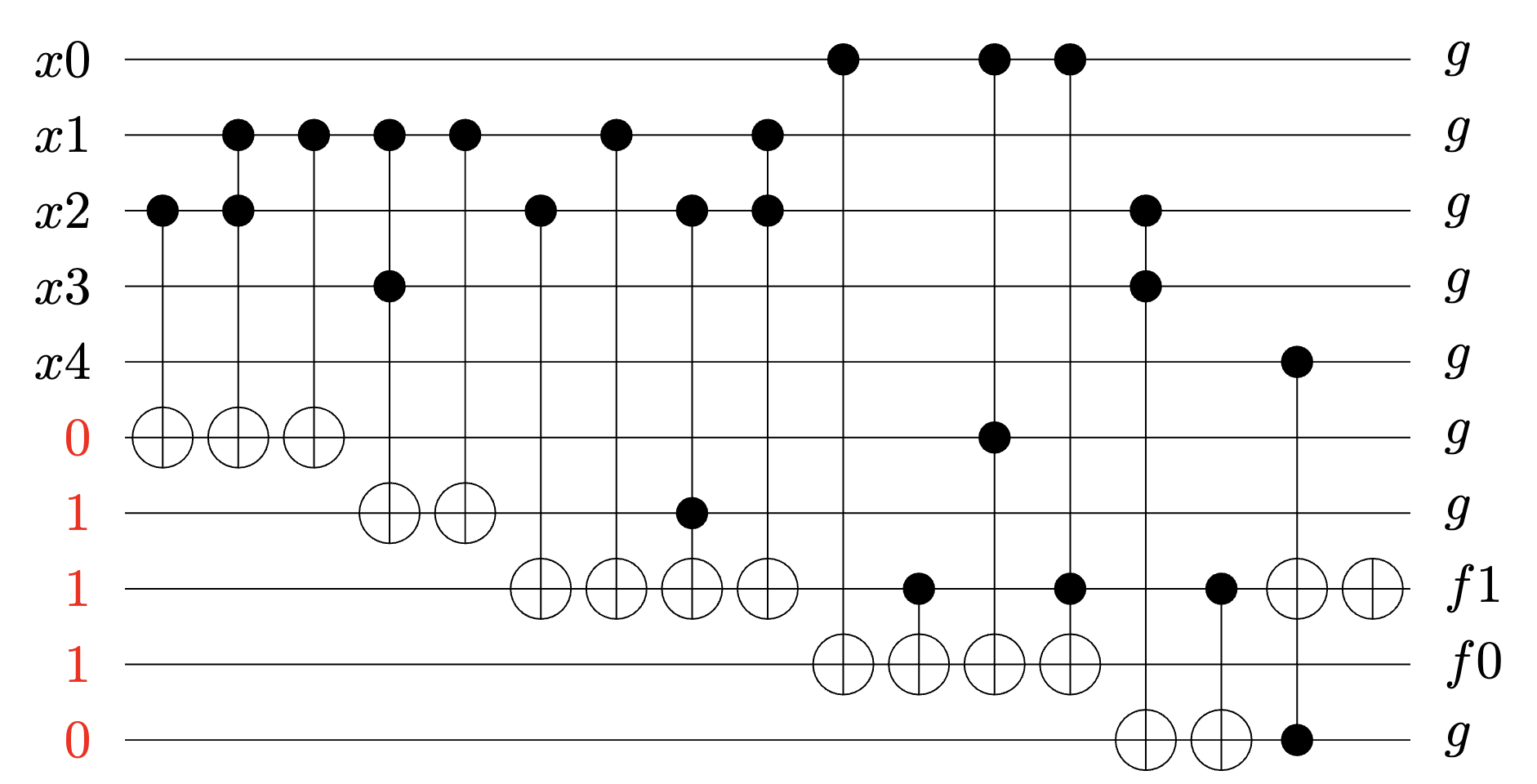} % Your first image
        \caption{Without Reordering}
        \label{c17_rev}
    \end{subfigure}%
    \hspace{0.5cm}  % Adjust the space between the images
    \begin{subfigure}{0.2\textwidth}  % Set the width of the subfigure
        \centering
        \includegraphics[width=\linewidth]{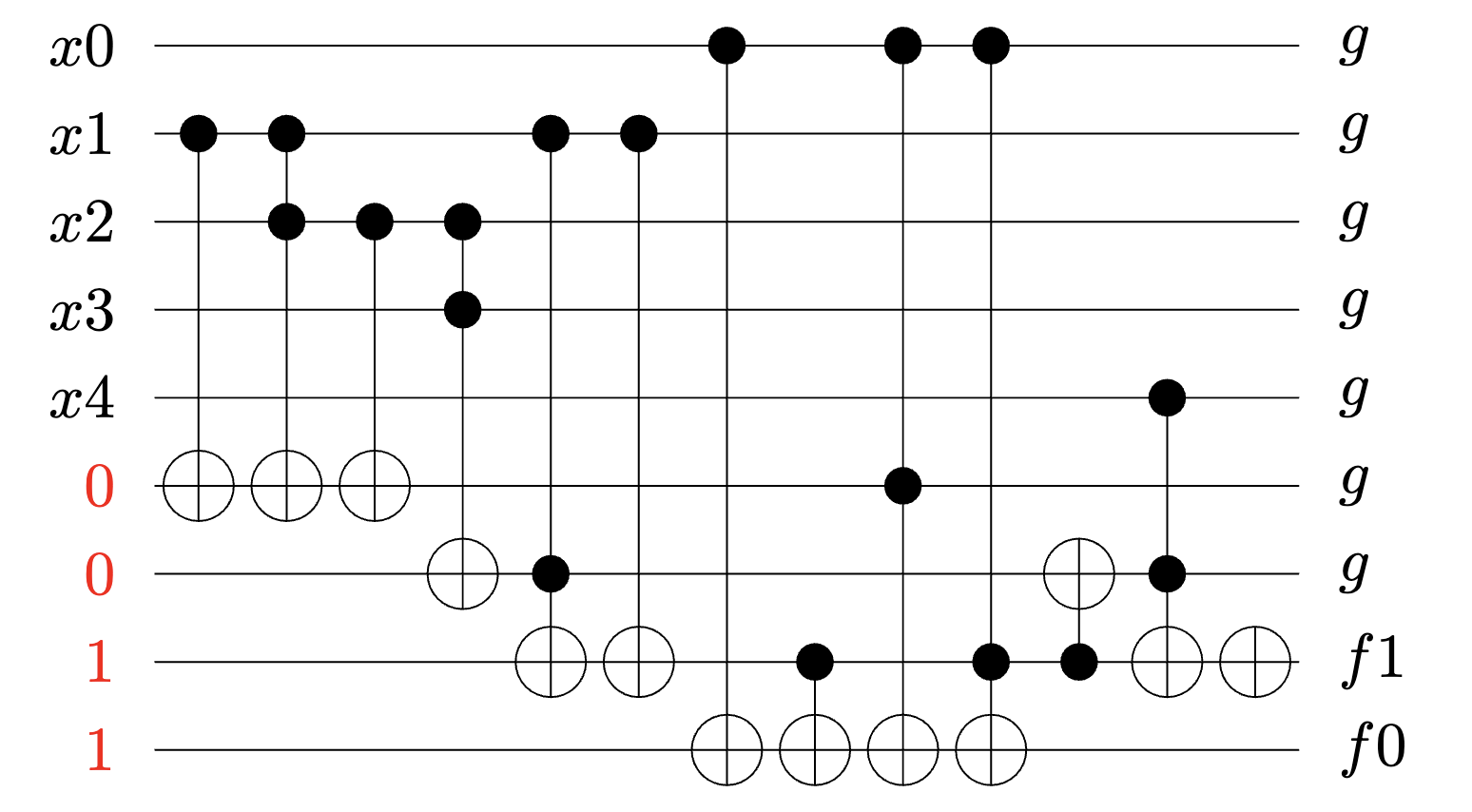} % Your second image
        \caption{With Reordering}
        \label{fig:second_image}
    \end{subfigure}
    \caption{Schematics for reversible C17 circuit w/wo GA}
    \label{c17_rev}
\end{figure}

\subsection{GNN Encoder Selection and Training Hyper-parameters}
To isolate the effect of the message-passing scheme itself, we evaluated four standard GNN encoders, namely, GCN, GIN, GraphSAGE, and GAT under identical architectural settings (hidden dimension $512$, $6$ layers).  
As summarised in \textbf{~\Cref{gnn_metrics}}, GAT attains the highest average Kendall’s~$\tau$ ($0.6182$) and Spearman’s~$\rho$ ($0.6748$); hence it is adopted as the default encoder in all subsequent experiments.

\begin{itemize}
  \item \textbf{GNN hidden dimension:} 512
  \item \textbf{Number of GNN layers:} 6
  \item \textbf{Batch size:} 16
  \item \textbf{Epochs:} 400
  \item \textbf{Optimizer:} Adam
  \item \textbf{Learning rate:} $1 \times 10^{-5}$
  \item \textbf{Random seed:} 42
\end{itemize}
\begin{table}[h!]
\centering
\small
\caption{Performance comparison for GNN encoders}
\label{gnn_metrics}
\begin{tabular}{@{}ccc@{}}
\toprule
\textbf{Encoder} & \textbf{Average Kendall's Tau} & \textbf{Average Spearman's Rho} \\ \midrule\midrule
GCN             & 0.4540                         & 0.5216                         \\
GIN             & 0.5129                         & 0.5877                         \\
GraphSage       & 0.5978                         & 0.6589                         \\
\textbf{GAT}             & \textbf{0.6182}                         & \textbf{0.6748}                         \\ \bottomrule 
\end{tabular}

\end{table}

\subsection{Comparison on BDD Nodes and Time Consumption}
\begin{figure}[!htbp] 
    \centering
    \includegraphics[width=1\linewidth]{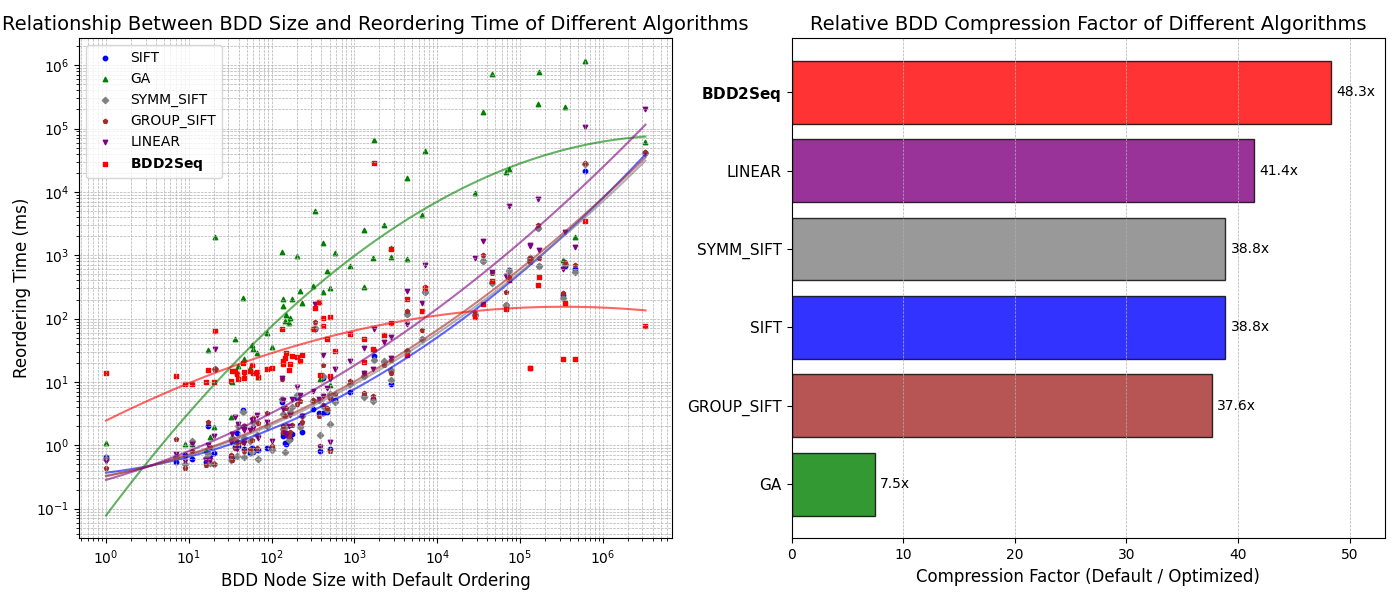}
    \caption{Time consumption and node size comparison}
    \label{CUDD_results}
\end{figure}

Having selected \textbf{GAT} as our encoder, we next evaluate our model's performance in terms of scalability and compression effectiveness compared with traditional heuristic algorithms on LGSynth91 benchmark. The comparative results are illustrated in \textbf{\Cref{CUDD_results}}, highlighting two key performance dimensions: time efficiency and node size reduction performance.

In terms of the time efficiency of reordering, the left side of \textbf{\Cref{CUDD_results}} compares the reordering time required by each algorithm relative to the original BDD node sizes before reordering. Quadratic polynomials  are fitted to the data points in a log-log scale to depict trends. Our approach consistently exhibits shorter and more predictable runtimes across varying node sizes compared to heuristic-based algorithms, whose runtimes grow substantially with node sizes. Heuristic-based algorithms show a rapid escalation in computation time as the circuit complexity grows, while our method maintains a relatively steady and efficient scaling trend, demonstrating better scalability.

On the right side of the \textbf{\Cref{CUDD_results}}, we quantify the performance of each algorithm through the relative BDD compression factor, defined as the ratio between original BDD node counts and optimized BDD node counts. Our method achieves the highest compression ratio \textbf{$(48.3\times)$}, outperforming all other traditional approaches. This result highlights the effectiveness of our model, indicating superior capability in learning optimal or near-optimal variable ordering compared to heuristic methods.

\subsection{Ablation study on \texttt{BLIF2Graph} and \texttt{AIG2Graph} on BDD variable ordering task}
To isolate the effect of the circuit--to--graph representation, we keep the graph encoder fixed as a GAT and hold all training/configuration settings constant (optimizer, budget, decoder, batching). The only change is the graph construction: our \texttt{BLIF2Graph} versus an \texttt{AIG2Graph} obtained by \texttt{strash} then edgelist extraction. On ISCAS85, AIG's 2-input normalization inflates graphs: average node and edge counts are \(2.40\times\) and \(1.39\times\) those of \texttt{BLIF2Graph}, respectively. Correspondingly, end-to-end wall-clock (preprocessing\,+\,GNN inference) is reduced by \(32.9\%\) with \texttt{BLIF2Graph}.

Despite being more compact, \texttt{BLIF2Graph} also improves order quality on the held-out test set: Kendall's \(\tau=0.6182\) and Spearman's \(\rho=0.6748\), versus \(\tau=0.6038\) and \(\rho=0.6517\) for \texttt{AIG2Graph}. For BDD variable ordering, preserving gate-functional semantics at the BLIF level yields smaller graphs, lower wall-clock, and better ranking accuracy than first reducing to AIG.

\subsection{Complete Comparison Results for Table 4}
\textbf{Next two pages} present the table for all successfully synthesized reversible circuits in evaluation using different algorithms. Due to its length, we have to split it into two separate tables across two pages.

        \begin{table*}
    \centering
    \caption{Quantum Cost and time consumption: \texttt{BDD2Seq} vs. traditional algorithms (Part 1)}
    \label{tab:all-results1}
    \resizebox{\linewidth}{!}{%
    \begin{tabular}{l|rr|rr|rr|rr|rr|rr|rc}
    \hline
    \multicolumn{1}{c|}{\multirow{2}{*}{Circuit}} &
    \multicolumn{2}{c|}{\texttt{BDD2Seq(B$^*$)}} & 
    \multicolumn{2}{c|}{\texttt{BDD2Seq(E$^*$)}} & 
    \multicolumn{2}{c|}{SIFT} & 
    \multicolumn{2}{c|}{SYMM\_SIFT} & 
    \multicolumn{2}{c|}{GROUP\_SIFT} & 
    \multicolumn{2}{c|}{GA} & 
    \multicolumn{2}{c}{LINEAR} \\
    \cline{2-15} 
     &
    \multicolumn{1}{c}{\small{QC$^\dag$}} & 
    \multicolumn{1}{c|}{\makecell[r]{\small{Time($s$)$^\ddag$}}} & 
    \multicolumn{1}{c}{\small{QC}} & 
    \multicolumn{1}{c|}{\makecell[r]{\small{Time($s$)}}} & 
    \multicolumn{1}{c}{\small{QC}} & 
    \multicolumn{1}{c|}{\makecell[r]{\small{Time($s$)}}} & 
    \multicolumn{1}{c}{\small{QC}} & 
    \multicolumn{1}{c|}{\makecell[r]{\small{Time($s$)}}} & 
    \multicolumn{1}{c}{\small{QC}} & 
    \multicolumn{1}{c|}{\makecell[r]{\small{Time($s$)}}} & 
    \multicolumn{1}{c}{\small{QC}} & 
    \multicolumn{1}{c|}{\makecell[r]{\small{Time($s$)}}} &
    \multicolumn{1}{c}{\small{QC}} & 
    \multicolumn{1}{c}{\makecell[r]{\small{Time($s$)}}} \\
    \hline

        \texttt{4gt10\_22.pla} &17& 1.22 &17& 2.13 &17& {0.01} &17& 0.07 &23& 0.05 &17& {0.01} &17& 0.05 \\ 
    \texttt{4gt11\_23.pla} &5& 0.07 &5& {0.01} &5& 0.01 &5& 0.04 &5& 0.06 &5& 0.01 &5& 0.06 \\ 
    \texttt{4gt12\_24.pla} &17& 0.06 &17& {0.01} &17& 0.01 &17& 0.05 &16& 0.05 &17& 0.01 &17& 0.05 \\ 
    \texttt{4gt13\_25.pla} &10& 0.07 &10& {0.01} &10& 0.01 &10& 0.05 &10& 0.05 &10& 0.01 &10& 0.05 \\ 
    \texttt{4gt4\_20.pla} &{19}& 0.06 &25& 0.01 &19& 0.01 &19& 0.06 &23& 0.06 &19& 0.01 &19& 0.05 \\ 
    \texttt{4gt5\_21.pla} &{12}& 0.07 &12& 0.01 &12& 0.01 &12& 0.06 &18& 0.05 &12& 0.01 &12& 0.05 \\ 
    \texttt{4mod5\_8.pla} &{24}& 0.07 &24& 0.02 &24& 0.01 &24& 0.05 &24& 0.06 &24& 0.01 &7& 0.05 \\ 
    \texttt{4mod7\_26.pla} &{86}& 0.06 &86& 0.01 &86& 0.01 &86& 0.06 &80& 0.05 &86& 0.01 &64& 0.05 \\ 
    \texttt{5xp1\_90.pla} &{254}& 0.13 &280& 0.01 &254& 0.01 &254& 0.07 &254& 0.06 &280& 0.01 &254& 0.06 \\ 
    \texttt{9symml\_91.pla} &{206}& 0.2 &206& 0.03 &206& 0.01 &206& 0.07 &206& 0.06 &206& 0.01 &206& 0.06 \\ 
    \texttt{C17\_117.pla} &{37}& 0.08 &37& 0.01 &49& 0.01 &37& 0.06 &37& 0.05 &37& 0.01 &49& 0.05 \\ 
    \texttt{C7552\_119.pla} &{202}& 0.08 &202& 0.01 &202& 0.01 &202& 0.06 &202& 0.05 &202& 0.01 &202& 0.06 \\ 
    \texttt{add6\_92.pla} &{118}& 0.29 &566& 0.01 &499& 0.01 &118& 0.11 &474& 0.08 &118& 0.02 &499& 0.08 \\ 
    \texttt{adr4\_93.pla} &{74}& 0.16 &74& 0.02 &74& 0.01 &74& 0.06 &74& 0.05 &74& 0.01 &47& 0.06 \\ 
    \texttt{alu1\_94.pla} &{139}& 0.29 &139& 0.02 &139& 0.01 &139& 0.07 &139& 0.09 &139& 0.01 &139& 0.05 \\ 
    \texttt{alu2\_96.pla} &{1218}& 0.23 &1415& 0.03 &1436& 0.01 &1299& 0.15 &1366& 0.08 &1298& 0.03 &1376& 0.08 \\ 
    \texttt{alu3\_97.pla} &{430}& 0.22 &464& 0.01 &644& 0.01 &430& 0.1 &497& 0.06 &430& 0.02 &644& 0.06 \\ 
    \texttt{alu4\_98.pla} &{4455}& 0.44 &4934& 0.09 &7222& 0.03 &4334& 0.65 &7623& 0.2 &4403& 0.18 &5583& 0.17 \\ 
    \texttt{alu\_9.pla} &{29}& 0.08 &29& 0.02 &29& 0.01 &29& 0.06 &35& 0.05 &29& 0.01 &29& 0.05 \\ 
    \texttt{apex2\_101.pla} &{3103}& 2.37 &3178& 0.16 &5922& 0.14 &2714& 1.57 &6439& 0.36 &2939& 2.78 &5922& 0.35 \\ 
    \texttt{apex4\_103.pla} &{8409}& 0.42 &8813& 0.19 &8343& 0.02 &8236& 0.55 &8381& 0.16 &8231& 0.07 &8336& 0.16 \\ 
    \texttt{apex5\_104.pla} &{9852}& 18.27 &10697& 0.2 &10349& 0.08 &9467& 3.62 &10227& 0.41 &10283& 1.47 &10349& 0.42 \\ 
    \texttt{apex6\_orig.pla} &{3895}& 23.86 &4021& 0.16 &4827& 0.26 &4815& 0.26 &4274& 0.26 &4046& 0.7 &4929& 0.06 \\ 
    \texttt{apex7\_orig.pla} &{1455}& 3.41 &1456& 0.07 &2652& 0.14 &2652& 0.13 &2377& 0.14 &1442& 0.48 &2215& 0.02 \\ 
    \texttt{apla\_107.pla} &{732}& 0.22 &955& 0.03 &1002& 0.01 &737& 0.11 &979& 0.07 &731& 0.01 &979& 0.06 \\ 
    \texttt{b1\_orig.pla} &{22}& 0.05 &22& 0.01 &22& 0.05 &22& 0.05 &22& 0.05 &22& 0.01 &10& 0.01 \\ 
    \texttt{b9\_orig.pla} &{737}& 2.43 &737& 0.05 &795& 0.09 &801& 0.08 &831& 0.09 &720& 0.2 &814& 0.01 \\ 
    \texttt{bw\_116.pla} &{924}& 0.08 &937& 0.02 &943& 0.01 &937& 0.09 &943& 0.07 &937& 0.01 &943& 0.06 \\ 
    \texttt{c17.pla} &{37}& 0.08 &49& 0.01 &49& 0.07 &49& 0.05 &37& 0.06 &37& 0.01 &49& 0.01 \\ 
    \texttt{c432.pla} &{12255}& 11.39 &20391& 6.79 &11101& 7.0 &11101& 6.73 &11141& 6.89 &10977& 9.14 &11101& 1.37 \\ 
    \texttt{c880.pla} &{37802}& 26.22 &39992& 15.52 &61131& 47.71 &61131& 44.43 &78157& 45.2 &35122& 121.45 &61216& 16.86 \\ 
    \texttt{c8\_orig.pla} &{432}& 1.21 &436& 0.04 &445& 0.06 &445& 0.07 &486& 0.07 &444& 0.12 &445& 0.01 \\ 
    \texttt{cc\_orig.pla} &{272}& 0.72 &278& 0.03 &379& 0.06 &379& 0.06 &379& 0.07 &287& 0.03 &379& 0.01 \\ 
    \texttt{cht\_orig.pla} &{714}& 3.15 &714& 0.06 &714& 0.07 &714& 0.07 &714& 0.07 &714& 0.09 &714& 0.01 \\ 
    \texttt{clip\_124.pla} &{695}& 0.18 &698& 0.02 &704& 0.01 &526& 0.1 &520& 0.07 &512& 0.01 &419& 0.07 \\ 
    \texttt{cm138a\_orig.pla} &{104}& 0.11 &104& 0.01 &104& 0.05 &104& 0.06 &104& 0.05 &104& 0.01 &104& 0.01 \\ 
    \texttt{cm150a\_128.pla} &{136}& 0.84 &136& 0.24 &186& 0.25 &186& 0.25 &186& 0.19 &186& 0.33 &186& 0.32 \\ 
    \texttt{cm150a\_orig.pla} &{136}& 0.93 &136& 0.34 &186& 0.18 &186& 0.18 &186& 0.21 &186& 0.3 &186& 0.37 \\ 
    \texttt{cm151a\_129.pla} &{298}& 0.61 &298& 0.02 &298& 0.01 &298& 0.11 &298& 0.06 &298& 0.05 &298& 0.06 \\ 
    \texttt{cm151a\_orig.pla} &{70}& 0.29 &70& 0.02 &92& 0.06 &92& 0.05 &92& 0.05 &92& 0.02 &92& 0.01 \\ 
    \texttt{cm152a\_130.pla} &{62}& 0.25 &62& 0.02 &62& 0.01 &62& 0.07 &62& 0.06 &62& 0.02 &62& 0.05 \\ 
    \texttt{cm152a\_orig.pla} &{62}& 0.25 &81& 0.02 &62& 0.06 &62& 0.05 &62& 0.06 &62& 0.01 &62& 0.01 \\ 
    \texttt{cm162a\_orig.pla} &{222}& 0.36 &222& 0.03 &192& 0.05 &247& 0.05 &219& 0.06 &219& 0.03 &192& 0.01 \\ 
    \texttt{cm163a\_133.pla} &{244}& 0.45 &267& 0.02 &273& 0.01 &267& 0.08 &267& 0.06 &244& 0.02 &261& 0.07 \\ 
    \texttt{cm163a\_orig.pla} &{136}& 0.45 &136& 0.03 &124& 0.06 &124& 0.05 &137& 0.06 &124& 0.02 &203& 0.01 \\ 
    \texttt{cm42a\_125.pla} &{117}& 0.07 &117& 0.01 &117& 0.01 &117& 0.05 &117& 0.06 &117& 0.01 &117& 0.05 \\ 
    \texttt{cm42a\_orig.pla} &{117}& 0.05 &117& 0.02 &117& 0.05 &117& 0.06 &117& 0.06 &117& 0.01 &117& 0.01 \\ 
    \texttt{cm82a\_126.pla} &{44}& 0.07 &44& 0.01 &82& 0.01 &44& 0.05 &44& 0.04 &44& 0.01 &49& 0.05 \\ 
    \texttt{cm82a\_orig.pla} &{44}& 0.07 &44& 0.01 &82& 0.05 &44& 0.05 &44& 0.06 &44& 0.01 &49& 0.01 \\ 
    \texttt{cm85a\_127.pla} &{227}& 0.25 &283& 0.03 &275& 0.01 &218& 0.08 &221& 0.05 &207& 0.02 &127& 0.06 \\ 
    \texttt{cm85a\_orig.pla} &{225}& 0.25 &225& 0.02 &275& 0.05 &285& 0.06 &221& 0.06 &207& 0.01 &127& 0.01 \\ 
    \texttt{cmb\_134.pla} &{153}& 0.45 &153& 0.02 &153& 0.01 &153& 0.09 &153& 0.09 &153& 0.03 &153& 0.06 \\ 
    \texttt{cmb\_orig.pla} &{153}& 0.45 &153& 0.03 &153& 0.06 &153& 0.05 &153& 0.05 &153& 0.03 &153& 0.01 \\ 
    \texttt{co14\_135.pla} &{159}& 0.36 &159& 0.02 &159& 0.01 &159& 0.08 &159& 0.06 &159& 0.03 &159& 0.06 \\ 
    \texttt{comp\_orig.pla} &{824}& 45.14 &824& 28.75 &961& 92.68 &1071& 87.08 &884& 92.8 &824& 69.75 &436& 35.81 \\ 
    \texttt{con1\_136.pla} &{88}& 0.13 &103& 0.03 &96& 0.01 &95& 0.06 &96& 0.05 &95& 0.01 &96& 0.05 \\ 
    \texttt{cordic\_138.pla} &{344}& 0.87 &446& 0.05 &325& 0.03 &325& 0.25 &315& 0.15 &306& 0.14 &220& 0.14 \\ 
    \texttt{cordic\_orig.pla} &{320}& 0.86 &320& 0.06 &325& 0.15 &323& 0.14 &315& 0.16 &306& 0.15 &220& 0.03 \\ 
    \texttt{count\_orig.pla} &{441}& 1.8 &441& 0.04 &441& 0.08 &441& 0.08 &441& 0.09 &441& 0.2 &441& 0.01 \\ 
    \texttt{cu\_141.pla} &{219}& 0.36 &230& 0.02 &220& 0.01 &220& 0.09 &224& 0.06 &224& 0.03 &234& 0.06 \\ 
    \texttt{cu\_orig.pla} &{224}& 0.36 &231& 0.02 &220& 0.06 &220& 0.06 &224& 0.06 &224& 0.02 &234& 0.01 \\ 
    \texttt{dalu\_orig.pla} &{5297}& 67.32 &5652& 28.17 &31145& 182.95 &31145& 212.63 &47170& 202.55 &5235& 270.63 &14330& 431.41 \\ 
    \texttt{dc1\_142.pla} &{160}& 0.07 &168& 0.01 &160& 0.01 &186& 0.05 &186& 0.06 &186& 0.01 &181& 0.05 \\ 
    \texttt{dc2\_143.pla} &{431}& 0.15 &431& 0.03 &431& 0.01 &431& 0.07 &431& 0.06 &431& 0.01 &431& 0.05 \\ 
    \texttt{decod24-enable\_32.pla} &{38}& 0.05 &38& 0.01 &38& 0.01 &38& 0.04 &38& 0.06 &38& 0.01 &38& 0.06 \\ 
    \texttt{decod\_137.pla} &{202}& 0.08 &202& 0.01 &202& 0.01 &202& 0.07 &202& 0.06 &202& 0.01 &202& 0.05 \\ 
    \texttt{decod\_orig.pla} &{202}& 0.07 &202& 0.01 &202& 0.06 &202& 0.06 &202& 0.06 &202& 0.01 &202& 0.01 \\ 
    \texttt{dist\_144.pla} &{975}& 0.17 &979& 0.04 &975& 0.01 &975& 0.12 &979& 0.08 &979& 0.02 &968& 0.08 \\ 
    \texttt{dk17\_145.pla} &{429}& 0.21 &429& 0.03 &426& 0.01 &426& 0.09 &574& 0.07 &426& 0.01 &426& 0.06 \\ 
    \texttt{dk27\_146.pla} &{131}& 0.18 &131& 0.02 &144& 0.01 &144& 0.06 &150& 0.05 &140& 0.01 &144& 0.06 \\ 
    \texttt{e64\_149.pla} &{1019}& 5.9 &1208& 0.17 &907& 0.02 &886& 1.8 &1124& 0.11 &886& 0.98 &907& 0.1 \\ 
    \hline   
    \end{tabular}
    }
    \begin{tablenotes}
        \small
        \item[*] * The letter in parentheses indicates synthesis mode, i.e., \texttt{(B)} means \textbf{{Balance}} mode, \texttt{(E)} means \textbf{{Efficiency}} mode.
        \item[\dag] \dag Quantum Cost 
        \item[\ddag] \ddag Synthesis times reported as 0.01$s$ represent Revkit's measurement limit ($\leq 0.01s$).
    \end{tablenotes}
    \end{table*}

    \begin{table*}
    \centering
    \caption{Quantum Cost and time consumption: \texttt{BDD2Seq} vs. traditional algorithms (Part 2)}
    \label{tab:all-results2}
    \resizebox{\linewidth}{!}{%
    \begin{tabular}{l|rr|rr|rr|rr|rr|rr|rc}
    \hline
    \multicolumn{1}{c|}{\multirow{2}{*}{Circuit}} &
    \multicolumn{2}{c|}{\texttt{BDD2Seq(B$^*$)}} & 
    \multicolumn{2}{c|}{\texttt{BDD2Seq(E$^*$)}} & 
    \multicolumn{2}{c|}{SIFT} & 
    \multicolumn{2}{c|}{SYMM\_SIFT} & 
    \multicolumn{2}{c|}{GROUP\_SIFT} & 
    \multicolumn{2}{c|}{GA} & 
    \multicolumn{2}{c}{LINEAR} \\
    \cline{2-15} 
     &
    \multicolumn{1}{c}{\small{QC$^\dag$}} & 
    \multicolumn{1}{c|}{\makecell[r]{\small{Time($s$)$^\ddag$}}} & 
    \multicolumn{1}{c}{\small{QC}} & 
    \multicolumn{1}{c|}{\makecell[r]{\small{Time($s$)}}} & 
    \multicolumn{1}{c}{\small{QC}} & 
    \multicolumn{1}{c|}{\makecell[r]{\small{Time($s$)}}} & 
    \multicolumn{1}{c}{\small{QC}} & 
    \multicolumn{1}{c|}{\makecell[r]{\small{Time($s$)}}} & 
    \multicolumn{1}{c}{\small{QC}} & 
    \multicolumn{1}{c|}{\makecell[r]{\small{Time($s$)}}} & 
    \multicolumn{1}{c}{\small{QC}} & 
    \multicolumn{1}{c|}{\makecell[r]{\small{Time($s$)}}} &
    \multicolumn{1}{c}{\small{QC}} & 
    \multicolumn{1}{c}{\makecell[r]{\small{Time($s$)}}} \\
    \hline
\texttt{ex1010\_155.pla} &{9633}& 0.41 &9810& 0.14 &9766& 0.03 &9824& 0.63 &9696& 0.21 &9792& 0.19 &9732& 0.21 \\ 
    \texttt{ex1\_150.pla} &{8}& 0.1 &8& 0.01 &8& 0.01 &8& 0.05 &8& 0.05 &8& 0.01 &8& 0.04 \\ 
    \texttt{ex2\_151.pla} &{60}& 0.09 &68& 0.01 &73& 0.01 &70& 0.05 &60& 0.06 &60& 0.01 &49& 0.05 \\ 
    \texttt{ex3\_152.pla} &{38}& 0.1 &38& 0.02 &61& 0.01 &48& 0.05 &43& 0.05 &48& 0.01 &37& 0.05 \\ 
    \texttt{ex5p\_154.pla} &{1837}& 0.22 &1871& 0.06 &1843& 0.01 &1843& 0.17 &1843& 0.11 &1843& 0.02 &1843& 0.1 \\ 
    \texttt{example2\_156.pla} &{1218}& 0.24 &1415& 0.03 &1436& 0.01 &1299& 0.15 &1366& 0.08 &1298& 0.02 &1376& 0.08 \\ 
    \texttt{example2\_orig.pla} &{1800}& 9.74 &1806& 0.08 &2085& 0.14 &2085& 0.15 &2063& 0.15 &1791& 0.53 &2000& 0.03 \\ 
    \texttt{f2\_158.pla} &{108}& 0.07 &113& 0.01 &113& 0.01 &113& 0.05 &113& 0.05 &108& 0.01 &100& 0.05 \\ 
    \texttt{f51m\_159.pla} &{2427}& 0.6 &2517& 0.14 &5392& 0.04 &2433& 0.49 &4603& 0.24 &2357& 0.16 &4825& 0.2 \\ 
    \texttt{frg1\_160.pla} &{598}& 1.34 &629& 0.03 &747& 0.01 &599& 0.25 &827& 0.08 &653& 0.2 &747& 0.08 \\ 
    \texttt{frg1\_orig.pla} &{757}& 1.35 &762& 0.03 &747& 0.07 &747& 0.07 &827& 0.07 &653& 0.2 &747& 0.01 \\ 
    \texttt{frg2\_161.pla} &{8584}& 26.96 &10768& 0.37 &12468& 0.25 &7313& 6.66 &12154& 1.36 &9391& 2.39 &12111& 1.34 \\ 
    \texttt{frg2\_orig.pla} &{7189}& 27.54 &7189& 0.35 &12468& 1.19 &12361& 1.2 &12154& 1.18 &9391& 2.41 &12111& 0.25 \\ 
    \texttt{in0\_162.pla} &{2296}& 0.42 &2417& 0.04 &2283& 0.02 &2327& 0.3 &2328& 0.09 &2303& 0.09 &2249& 0.09 \\ 
    \texttt{inc\_170.pla} &{579}& 0.13 &579& 0.01 &579& 0.01 &592& 0.08 &592& 0.07 &592& 0.01 &621& 0.06 \\ 
    \texttt{k2\_orig.pla} &{10432}& 3.33 &10686& 0.19 &11058& 0.29 &11058& 0.29 &11302& 0.31 &10304& 2.08 &10894& 0.11 \\ 
    \texttt{lal\_orig.pla} &{478}& 1.07 &496& 0.04 &527& 0.06 &774& 0.06 &724& 0.07 &506& 0.08 &464& 0.01 \\ 
    \texttt{life\_175.pla} &{204}& 0.19 &210& 0.01 &204& 0.01 &204& 0.09 &204& 0.08 &210& 0.02 &179& 0.07 \\ 
    \texttt{majority\_176.pla} &{41}& 0.08 &41& 0.01 &41& 0.01 &41& 0.05 &41& 0.05 &41& 0.01 &41& 0.05 \\ 
    \texttt{majority\_orig.pla} &{41}& 0.07 &41& 0.01 &41& 0.05 &41& 0.06 &41& 0.05 &41& 0.01 &41& 0.01 \\ 
    \texttt{max46\_177.pla} &{562}& 0.19 &596& 0.02 &598& 0.01 &561& 0.08 &598& 0.06 &561& 0.01 &598& 0.06 \\ 
    \texttt{misex1\_178.pla} &{281}& 0.16 &289& 0.02 &304& 0.01 &279& 0.07 &302& 0.06 &287& 0.01 &304& 0.06 \\ 
    \texttt{misex3\_180.pla} &{3830}& 0.44 &4054& 0.09 &4661& 0.04 &3795& 0.6 &4619& 0.24 &3789& 0.21 &4657& 0.24 \\ 
    \texttt{misex3c\_181.pla} &{3969}& 0.4 &4457& 0.05 &4769& 0.02 &4165& 0.57 &4823& 0.14 &3945& 0.23 &4725& 0.14 \\ 
    \texttt{mlp4\_184.pla} &{1191}& 0.17 &1228& 0.03 &1158& 0.01 &1158& 0.11 &1158& 0.07 &1159& 0.02 &1158& 0.06 \\ 
    \texttt{mux\_185.pla} &{135}& 1.18 &224& 0.26 &170& 0.59 &170& 0.34 &170& 0.28 &170& 0.63 &170& 0.39 \\ 
    \texttt{mux\_orig.pla} &{135}& 1.03 &135& 0.4 &170& 0.24 &170& 0.23 &170& 0.26 &170& 0.51 &170& 0.43 \\ 
    \texttt{my\_adder\_orig.pla} &{352}& 64.22 &352& 48.54 &352& 169.32 &352& 171.76 &352& 165.44 &352& 74.96 &208& 58.36 \\ 
    \texttt{one-two-three\_27.pla} &{44}& 0.05 &44& 0.01 &44& 0.01 &44& 0.05 &44& 0.05 &44& 0.01 &44& 0.05 \\ 
    \texttt{pair\_orig.pla} &{20754}& 39.87 &20799& 1.06 &46544& 4.86 &46444& 4.81 &49917& 4.91 &1039245& 18.63 &47818& 1.03 \\ 
    \texttt{parity\_188.pla} &{31}& 0.99 &31& 0.44 &31& 0.55 &31& 2.08 &31& 2.02 &31& 0.59 &31& 2.01 \\ 
    \texttt{parity\_orig.pla} &{31}& 0.92 &31& 0.61 &31& 2.1 &31& 2.07 &31& 2.13 &31& 0.62 &31& 0.55 \\ 
    \texttt{pcle\_orig.pla} &{298}& 0.61 &298& 0.02 &298& 0.06 &310& 0.05 &298& 0.05 &298& 0.05 &298& 0.01 \\ 
    \texttt{pcler8\_190.pla} &{124}& 0.46 &137& 0.01 &124& 0.01 &124& 0.08 &137& 0.06 &124& 0.02 &203& 0.06 \\ 
    \texttt{pcler8\_orig.pla} &{765}& 1.13 &765& 0.02 &639& 0.06 &639& 0.06 &639& 0.07 &639& 0.13 &639& 0.01 \\ 
    \texttt{pdc\_191.pla} &{6742}& 0.72 &6742& 0.23 &6500& 0.09 &6599& 1.19 &6500& 0.47 &6599& 0.39 &6098& 0.42 \\ 
    \texttt{pm1\_192.pla} &{117}& 0.06 &117& 0.02 &117& 0.01 &117& 0.06 &117& 0.06 &117& 0.01 &117& 0.06 \\ 
    \texttt{pm1\_orig.pla} &{234}& 0.46 &273& 0.03 &273& 0.06 &244& 0.05 &267& 0.06 &244& 0.02 &261& 0.01 \\ 
    \texttt{radd\_193.pla} &{74}& 0.16 &140& 0.01 &217& 0.01 &74& 0.07 &212& 0.07 &74& 0.01 &217& 0.06 \\ 
    \texttt{rd53\_68.pla} &{98}& 0.07 &98& 0.01 &98& 0.01 &98& 0.06 &98& 0.05 &98& 0.01 &65& 0.07 \\ 
    \texttt{rd73\_69.pla} &{217}& 0.12 &217& 0.02 &217& 0.01 &217& 0.06 &217& 0.06 &217& 0.01 &148& 0.06 \\ 
    \texttt{rd84\_70.pla} &{304}& 0.17 &304& 0.02 &304& 0.01 &304& 0.07 &304& 0.07 &304& 0.01 &233& 0.07 \\ 
    \texttt{root\_197.pla} &{444}& 0.16 &446& 0.01 &444& 0.01 &444& 0.08 &444& 0.06 &446& 0.01 &452& 0.06 \\ 
    \texttt{rot\_orig.pla} &{38453}& 47.58 &48344& 15.69 &78639& 61.42 &78375& 60.18 &110887& 61.68 &7339159& 139.15 &70456& 20.52 \\ 
    \texttt{ryy6\_198.pla} &{103}& 0.46 &107& 0.02 &133& 0.01 &107& 0.09 &119& 0.06 &119& 0.04 &133& 0.05 \\ 
    \texttt{s1196\_orig.pla} &{4104}& 0.41 &4110& 0.05 &4691& 0.19 &4691& 0.18 &4653& 0.18 &3765& 0.17 &4688& 0.04 \\ 
    \texttt{sao2\_199.pla} &{657}& 0.22 &698& 0.01 &667& 0.01 &653& 0.11 &684& 0.06 &657& 0.02 &716& 0.07 \\ 
    \texttt{sct\_orig.pla} &{405}& 0.61 &513& 0.02 &545& 0.06 &545& 0.07 &548& 0.07 &387& 0.04 &539& 0.01 \\ 
    \texttt{seq\_201.pla} &{9349}& 2.89 &9908& 0.33 &19362& 0.43 &9216& 3.54 &15309& 0.7 &9113& 4.17 &15364& 1.14 \\ 
    \texttt{sf\_232.pla} &{31}& 0.07 &36& 0.02 &36& 0.01 &36& 0.05 &21& 0.05 &43& 0.01 &36& 0.06 \\ 
    \texttt{spla\_202.pla} &{5947}& 0.66 &6092& 0.19 &5925& 0.07 &5858& 0.88 &5925& 0.36 &5858& 0.31 &5648& 0.36 \\ 
    \texttt{sqn\_203.pla} &{374}& 0.13 &426& 0.02 &426& 0.01 &361& 0.08 &426& 0.06 &356& 0.01 &392& 0.06 \\ 
    \texttt{sqr6\_204.pla} &{470}& 0.11 &524& 0.02 &486& 0.01 &486& 0.07 &486& 0.06 &482& 0.01 &486& 0.06 \\ 
    \texttt{sqrt8\_205.pla} &{237}& 0.16 &283& 0.01 &240& 0.01 &237& 0.07 &240& 0.06 &221& 0.01 &240& 0.06 \\ 
    \texttt{squar5\_206.pla} &{232}& 0.09 &264& 0.02 &253& 0.01 &232& 0.07 &232& 0.06 &232& 0.01 &253& 0.07 \\ 
    \texttt{sym10\_207.pla} &{253}& 0.24 &253& 0.03 &253& 0.01 &253& 0.1 &253& 0.09 &253& 0.02 &253& 0.1 \\ 
    \texttt{sym6\_63.pla} &{93}& 0.11 &93& 0.01 &93& 0.01 &93& 0.06 &93& 0.06 &93& 0.01 &93& 0.05 \\ 
    \texttt{sym9\_71.pla} &{206}& 0.2 &206& 0.03 &206& 0.01 &206& 0.07 &206& 0.06 &206& 0.01 &206& 0.06 \\ 
    \texttt{t481\_208.pla} &{139}& 0.47 &140& 0.03 &152& 0.01 &152& 0.11 &152& 0.08 &152& 0.04 &152& 0.08 \\ 
    \texttt{table3\_209.pla} &{5954}& 0.48 &7005& 0.08 &6276& 0.01 &5905& 0.76 &5927& 0.13 &5905& 0.23 &6294& 0.13 \\ 
    \texttt{tcon\_orig.pla} &{88}& 0.51 &88& 0.02 &88& 0.05 &88& 0.06 &88& 0.06 &88& 0.01 &88& 0.01 \\ 
    \texttt{term1\_orig.pla} &{480}& 1.76 &483& 0.06 &1193& 0.08 &1151& 0.09 &1070& 0.08 &503& 0.26 &1154& 0.01 \\ 
    \texttt{tial\_214.pla} &{4475}& 0.43 &5157& 0.09 &7609& 0.03 &4284& 0.59 &4852& 0.14 &4284& 0.18 &6952& 0.15 \\ 
    \texttt{too\_large\_orig.pla} &{3744}& 2.93 &5379& 0.46 &5922& 0.25 &5393& 0.25 &6334& 0.26 &2767& 2.51 &5922& 0.08 \\ 
    \texttt{ttt2\_orig.pla} &{749}& 0.92 &1006& 0.04 &727& 0.07 &727& 0.07 &1135& 0.08 &734& 0.1 &722& 0.01 \\ 
    \texttt{unreg\_orig.pla} &{494}& 1.93 &494& 0.04 &494& 0.06 &494& 0.06 &494& 0.06 &495& 0.1 &494& 0.01 \\ 
    \texttt{urf4\_89.pla} &{28488}& 0.77 &28833& 0.47 &28523& 0.09 &28733& 1.83 &28523& 0.44 &28393& 0.64 &28395& 0.45 \\ 
    \texttt{vda\_orig.pla} &{4224}& 0.56 &4224& 0.08 &4477& 0.14 &4477& 0.14 &4460& 0.15 &4169& 0.19 &4470& 0.03 \\ 
    \texttt{wim\_220.pla} &{103}& 0.07 &108& 0.01 &107& 0.01 &107& 0.05 &107& 0.05 &107& 0.01 &93& 0.05 \\ 
    \texttt{x1\_orig.pla} &{3567}& 3.76 &3703& 0.16 &3760& 0.13 &4277& 0.13 &4203& 0.14 &3172& 0.9 &3760& 0.03 \\ 
    \texttt{x2\_223.pla} &{191}& 0.23 &191& 0.03 &274& 0.01 &191& 0.08 &283& 0.06 &191& 0.01 &257& 0.05 \\ 
    \texttt{x2\_orig.pla} &{191}& 0.2 &191& 0.02 &274& 0.05 &274& 0.06 &283& 0.07 &191& 0.01 &257& 0.01 \\ 
    \texttt{x3\_orig.pla} &{3941}& 23.93 &4645& 0.19 &4827& 0.26 &4815& 0.25 &4274& 0.28 &4046& 0.7 &4929& 0.04 \\ 
    \texttt{x4\_orig.pla} &{2662}& 11.84 &2782& 0.11 &4459& 0.19 &4459& 0.2 &4334& 0.2 &3006& 0.64 &4459& 0.04 \\ 
    \texttt{xor5\_195.pla} &{8}& 0.07 &8& 0.01 &8& 0.01 &8& 0.05 &8& 0.05 &8& 0.01 &8& 0.06 \\ 
    \texttt{z4\_224.pla} &{66}& 0.13 &66& 0.01 &66& 0.01 &66& 0.06 &66& 0.05 &66& 0.01 &39& 0.05 \\ 
    \texttt{z4ml\_225.pla} &{66}& 0.13 &66& 0.02 &66& 0.01 &66& 0.06 &66& 0.08 &66& 0.01 &39& 0.05 \\
    \hline   
    \end{tabular}
    }
    \begin{tablenotes}
        \small
        \item[*] * The letter in parentheses indicates synthesis mode, i.e., \texttt{(B)} means \textbf{{Balance}} mode, \texttt{(E)} means \textbf{{Efficiency}} mode.
        \item[\dag] \dag Quantum Cost 
        \item[\ddag] \ddag Synthesis times reported as 0.01$s$ represent Revkit's measurement limit ($\leq 0.01s$).
    \end{tablenotes}
    \end{table*}

\end{document}